\documentclass [12pt,eqsecnum,amsfonts,aps]{revtex4}

\input epsf
\newcommand{\beq}{\begin{equation}}
\newcommand{\eeq}{\end{equation}}
\newcommand{\beqs}{\begin{eqnarray}}
\newcommand{\eeqs}{\end{eqnarray}}

\begin{document}

\baselineskip 6.0mm

\title{Weighted-Set Graph Colorings} 

\bigskip

\author{Robert Shrock}
\email{robert.shrock@stonybrook.edu}

\author{Yan Xu}
\email{yan.xu@stonybrook.edu}

\affiliation{ C. N. Yang Institute for Theoretical Physics \\
State University of New York \\
Stony Brook, N. Y. 11794}

%\bigskip

\begin{abstract}

We study a weighted-set graph coloring problem in which one assigns $q$ colors
to the vertices of a graph such that adjacent vertices have different colors,
with a vertex weighting $w$ that either disfavors or favors a given subset of
$s$ colors contained in the set of $q$ colors. We construct and analyze a
weighted-set chromatic polynomial $Ph(G,q,s,w)$ associated with this coloring.
General properties of this weighted-set chromatic polynomial are proved, and
illustrative calculations are presented for various families of graphs.  This
study extends a previous one for the case $s=1$ and reveals a number of
interesting new features.

\end{abstract}

\maketitle

%\newpage

\pagestyle{plain}
\pagenumbering{arabic}

\section{Introduction}

Recently, two weighted graph coloring problems have been formulated and studied
in which one assigns $q$ colors to the vertices of a graph subject to the
condition that adjacent vertices (i.e., vertices connected by an edge of the
graph) have different colors, with a vertex weighting $w$ that either disfavors
(for $0 \le w < 1$) or favors (for $w > 1$) a given color \cite{hl,ph}.  Since
all of the colors are, {\it a priori}, equivalent, it does not matter which
color one takes to be given the weighting.  An assignment of $q$ colors to the
vertices of a graph $G$, such that adjacent vertices have different colors, is
called a ``proper $q$-coloring'' of the vertices of $G$.  In the present paper
we shall study a generalization of this problem in which one performs a proper
$q$-coloring of the vertices of a graph $G$ such that $s$ colors are favored or
disfavored relative to the remaining $q-s$ colors.  We denote these coloring
problems as the DFSCP and FSCP for \underline{d}is\underline{f}avored or
\underline{f}avored weighted-\underline{s}et graph vertex \underline{c}oloring
\underline{p}roblems. We analyze the properties of an associated weighted-set
chromatic polynomial, denoted $Ph(G,q,s,w)$, which generalizes the chromatic
polynomial $P(G,q)$ and the single-color weighted chromatic polynomial
$Ph(G,q,w) \equiv Ph(G,q,1,w)$ analyzed in Ref. \cite{ph}.  We shall denote the
set of integers $\{1,...,s\}$, representing colors, as $I_s$ and the orthogonal
complement $\{s+1,...,q\}$ as $I_s^\perp$.  To each proper $q$-coloring of the
vertices of a graph $G$ there corresponds a term $w^{n_s}$, where $n_s$ denotes
the number of vertices assigned a color in $I_s$. The sum of such terms
resulting from all of these proper $q$-colorings of the vertices of $G$ is the
function $Ph(G,q,s,w)$.  As we shall show below, this is a polynomial not only
in $w$, but also in $q$ and $s$.  This polynomial constitutes a $w$-dependent
measure, extended from the integers to the real numbers, of the number of
proper $q$-colorings of the vertices of $G$.  In the weighted-set graph
coloring problem for a given graph $G$, with $q \in {\mathbb N}_+$ being the
number of colors, $Ph(G,q,s,w)$ is a map from $(q,s,w) \in {\mathbb N}_+ \times
I_s \times [0,\infty)$ to ${\mathbb R}$. One can formally extend the domain of
each of the variables $q$, $s$, and $w$ to ${\mathbb R}$ or, indeed, ${\mathbb
C}$, and the latter extension is necessary when one analyzes the zeros of
$Ph(G,q,s,w)$.  The polynomial $Ph(G,q,s,w)$ is equivalent to the partition
function of the $q$-state Potts antiferromagnet on the graph $G$ in a set of
external magnetic fields, in the limit where the effective spin-spin exchange
coupling becomes infinitely strong, so that the only spin configurations
contributing to this partition function are those for which spins on adjacent
vertices are different \cite{hl,ph,zth}. There has been continuing interest in
the Potts model and chromatic and Tutte polynomials for many years; reviews of
the Potts model include \cite{wurev}-\cite{wubook} and reviews of chromatic and
Tutte polynomials include \cite{rrev}-\cite{jemrev}.

There are several motivations for this study, arising from the areas of
mathematics, physics, and engineering.  One motivation is the intrinsic
mathematical interest in graph coloring problems and the fact that there seems
to have been very little previous study of weighted-set graph coloring.  A
second one stems from the equivalence to the statistical mechanics of the Potts
antiferromagnet in a set of magnetic fields that disfavor or favor a
corresponding set of spin values.  A third reason for interest in this subject
is the fact that these weighted-set graph coloring problems have practical
applications. For example, the weighted graph coloring problem with $0 \le w <
1$ (i.e., the DFSCP) describes, among other things, the assignment of
frequencies to commercial radio broadcasting stations in an area such that (i)
adjacent stations must use different frequencies to avoid interference and (ii)
stations prefer to avoid transmitting on a set of $s$ specific frequencies,
e.g., because these are used for data-taking by a nearby radio astronomy
antenna.  The weighted graph coloring problem with $w > 1$ (i.e., the FSCP)
describes this frequency assignment process with a preference for a set of $s$
frequencies, e.g., because these are most free of interference.  We shall
especially emphasize the connections with the first two of these areas in this
paper.

We note some special cases of the weighted-set chromatic polynomial.  Let us
consider a graph $G=(V,E)$, defined by its set of vertices $V$ and edges (=
bonds) $E$.  We denote the numbers of vertices and edges of $G$ as $n(G) \equiv
n$ and $e(G)$.  The values $s=0$ and $w=1$ correspond to the usual unweighted
proper $q$-coloring of the vertices of $G$, so $Ph(G,q,s,w)$ reduces to the
usual chromatic polynomial counting the number of proper $q$-colorings of the
vertices of $G$:
\beq
Ph(G,q,s,1) = Ph(G,q,0,w)=P(G,q) \ . 
\label{phw1andphs0}
\eeq
Since the right-hand side of Eq. (\ref{phw1andphs0}) is independent of $s$ and
$w$, this relation also implies the differential equations
\beq
\frac{\partial Ph(G,q,s,1)}{\partial s} = 0 
\label{dphw1ds}
\eeq
and
\beq
\frac{\partial Ph(G,q,0,w)}{\partial w} = 0 \ . 
\label{dphw1dw}
\eeq
For $w = 0$, one is prevented from assigning any of the $s$ disfavored colors
to any of the vertices, so that the problem reduces to that of a proper
coloring of the vertices of $G$ with $q-s$ colors, without any weighting among
them.  This is described by the usual (unweighted) chromatic polynomial
$P(G,q-s)$, so
\beq
Ph(G,q,s,0) = P(G,q-s) \ . 
\label{phw0}
\eeq
Thus, the DFSCP, described by $Ph(G,q,s,w)$ may be regarded as interpolating
between $P(G,q)$ and $P(G,q-s)$ as $w$ decreases through real values from $w=1$
to $w=0$. (The case of no weighting, $w=1$, may be considered to be the border
between the DFSCP and FSCP regimes.) If $s=q$, so that all of the colors
receive the same weighting, then, as is clear from its definition, the
weighted-set chromatic polynomial reduces to $w^n$ times the unweighted
chromatic polynomial:
\beq
Ph(G,q,q,w) = w^nP(G,q) \ . 
\label{phsq}
\eeq
Thus, for $s=0$ and $s=q$, $Ph(G,q,s,w)$ reduces to 1 and $w^n$ times $P(G,q)$,
respectively, while for other values of $s$, in particular, for integer $s$ in
the interval $1 \le s \le q-1$, $Ph(G,q,s,w)$ is a new polynomial which is not,
in general, reducible to $P(G,q)$.  Hence, while retaining the term DFSCP, we
shall often focus on the new cases where $w$ lies strictly between 1 and 0.  As
we shall show below, the weighted-set chromatic polynomial $Ph(G,q,s,w)$
satisfies a basic symmetry relation involving the interchange of $s$ with
$q-s$, so that a knowledge of the weighted-set proper $q$-coloring of the
vertices of a graph $G$ with a set of $s$ colors is equivalent to a knowledge
of the proper coloring of the vertices of $G$ with a set of $q-s$ colors.

There are important differences between the case $s=1$ studied previously in
Ref. \cite{ph} and the cases $2 \le s \le q$.  For $s=1$, as $w$ increases
above 1 to large positive values, the favored weighting of one color is
increasingly in conflict with the strict constraint that no two adjacent
vertices have the same color.  Hence, this involves competing interactions and
frustration. In contrast, in the FSCP regime with $s \ge 2$, depending on the
graph $G$, one may avoid this conflict and the resultant frustration.  Specific
differences will be apparent in our explicit results.  For example, in our
general result for $Z(G,q,s,v,w)$ for the circuit graph $C_n$ in
Eq. (\ref{zcnexplicit}) below, one term vanishes identically in the case $s=1$
but is present for other values of $s$ in the interval $I_s$.  (By the $s
\leftrightarrow q-s$ symmetry in Eqs. (\ref{symgroup}) and (\ref{zsym}), this
also means that another term vanishes identically for $s=q-1$ but is present
for other values of $s \in I_s$.)

\section{Some Basic Properties}

\subsection{Connection of $Ph(G,q,s,w)$ with Statistical Mechanics}

  It is useful to see how the function $Ph(G,q,s,w)$ arises in a more general
statistical mechanical context.  As before, we have $G=(V,E)$. A spanning
subgraph $G' \subseteq G$ is defined as the subgraph containing the same set of
vertices $V$ and a subset of the edges of $G$; $G' = (V,E')$ with $E' \subseteq
E$. We denote the number of connected components of $G$ as $k(G)$ and the
connected subgraphs of a spanning subgraph $G'$ as $G'_i$, $i=1,..,k(G')$.  To
obtain an expression for $Ph(G,q,s,w)$, we make use of the fact that it is a
special case of the partition function for the $q$-state Potts model in the
presence of external magnetic fields in the limit of infinitely strong
antiferromagnetic spin-spin coupling.  In thermal equilibrium at temperature
$T$, the general Potts model partition function is given by
\beq
Z = \sum_{ \{ \sigma_n \} } e^{-\beta {\cal H}} 
\label{z}
\eeq
with the Hamiltonian
\beq
{\cal H} = -J \sum_{\langle i j \rangle} \delta_{\sigma_i, \sigma_j}
- \sum_{p=1}^q H_p \sum_\ell \delta_{\sigma_\ell,p} \ , 
\label{ham}
\eeq
where $i, \ j, \ \ell$ label vertices of $G$, $\sigma_i$ are
classical spin variables on these vertices, taking values in the set 
$I_q = \{1,...,q\}$, $\beta = (k_BT)^{-1}$, $\langle
i j \rangle$ denote pairs of adjacent vertices, $p$ is an integer, 
$p \in I_q$, and $H_p$ is an external magnetic field given by 
\beq
H_p = \cases{ H \ne 0 & for $1 \le p \le s$ \cr
              0 & for $s+1 \le p \le q$ } \ . 
\label{hp}
\eeq
The zero-field Potts model Hamiltonian ${\cal H}$ and partition function $Z$
are invariant under the global transformation in which $\sigma_i \to g \sigma_i
\ \forall \ i \in V$, with $g \in S_q$, where $S_q$ is the symmetric (=
permutation) group on $q$ objects.  Because of this invariance, we can, without
loss of generality, take the external magnetic fields $H_p$ to single out a set
of $s$ contiguous spin values (equivalently, colors) $\sigma_i \in I_s$ as
disfavored or favored, relative to the orthogonal complement of values
$\sigma_i \in I_s^\perp$. In the presence of the magnetic fields $H_p$ given in
Eq. (\ref{hp}), the symmetry group of ${\cal H}$ and $Z$ is reduced to the
tensor product
\beq
S_q \to S_s \otimes S_{q-s} \ . 
\label{symgroup}
\eeq
That is, if $g_1 \in S_q$ and $g_2 \in S_{q-s}$, then the global transformation
$\sigma_i \to (g_1 \otimes g_2)\sigma_i \ \forall \ i$ leaves ${\cal H}$ and
$Z$ invariant. Here $(g_1 \otimes g_2)\sigma_i$ means $g_1 \sigma_i$ if
$\sigma_i \in I_s$ and $g_2 \sigma_i$ if $\sigma_i \in I_s^\perp$.  

Let us introduce the notation
\beq
K = \beta J \ , \quad h = \beta H \ , \quad y = e^K \ , \quad v = y-1 \ . 
\quad w=e^h \ . 
\label{kdef}
\eeq
Thus, the physical ranges of $v$ are $v \ge 0$ for the Potts ferromagnet, and
$-1 \le v \le 0$ for the Potts antiferromagnet. The weighted-set chromatic
polynomial is then obtained by choosing the antiferromagnetic sign of the
spin-spin coupling, $J < 0$ and taking $K \to -\infty$ while keeping $h=\beta
H$ fixed. Since $K=\beta J$, the limit $K \to -\infty$ results if (i) one takes
$J \to -\infty$ while holding $T$ and $H$ fixed and finite, or (ii) one takes
$T \to 0$, i.e., $\beta \to \infty$, with $J$ fixed and finite and $H \to 0$ so
as to keep $h = \beta H$ fixed and finite.  The limit $K \to -\infty$
guarantees that no two adjacent spins have the same value, or, in the coloring
context, that no two vertices have the same color.  One sees that in this
statistical mechanics context, it is the external magnetic fields that produce
the weighting that favors or disfavors a given set $I_s$ of spin values.
Positive $H$ gives a weighting that favors spin configurations in which spins
have values in the set $I_s$, or equivalently, vertex colorings with colors in
this set, while negative $H$ disfavors such configurations. For positive and
negative $H$, the corresponding ranges of $w$ are $w > 1$ and $0 \le w < 1$,
respectively.

In Ref. \cite{ph} a formula was derived for the partition function $Z$ which
does not make any explicit reference to the spins $\sigma_i$ or the summation
over spin configurations, but instead expresses this function as a sum of terms
arising from the $2^{e(G)}$ spanning subgraphs $G' \subseteq G$, namely 
\beq
Z(G,q,s,v,w) = \sum_{G' \subseteq G} v^{e(G')} \
\prod_{i=1}^{k(G')} \Big ( q-s + sw^{n(G'_i)} \Big ) \ . 
\label{clusterws}
\eeq
This generalizes a spanning subgraph formula for $Z$ in the case $s=1$ due to
F. Y. Wu \cite{wu78}, which, itself, generalized the Fortuin-Kasteleyn formula
for the zero-field Potts model partition function,
\beq
Z(G,q,v) = \sum_{G' \subseteq G} v^{e(G')} \ q^{k(G')} \ . 
\label{fk}
\eeq
The original definition of the Potts model, (\ref{z}) and (\ref{ham}), requires
$q$ to be in the set of positive integers ${\mathbb N}_+$ and $s$ to be a
non-negative integer. These restrictions are removed by Eq. (\ref{clusterws}).
Furthermore, Eq. (\ref{clusterws}) shows that $Z$ is a polynomial in the
variables $q$, $s$, $v$, and $w$, hence our notation $Z(G,q,s,v,w)$.

The $K \to -\infty$ limit that yields the weighted-set chromatic polynomial is
equivalent to $v=-1$, so
\beq
Ph(G,q,s,w) = Z(G,q,s,-1,w) \ . 
\label{phz}
\eeq
Hence, 
\beq
Ph(G,q,s,w) = \sum_{G' \subseteq G} (-1)^{e(G')} \
\prod_{i=1}^{k(G')} \Big ( q-s + sw^{n(G'_i)} \Big ) \ . 
\label{phclusterws}
\eeq
We recall the factorization
\beq
w^m-1=(w-1)\sum_{j=0}^{m-1}w^j
\label{wfactorformula}
\eeq
and apply it to Eq. (\ref{clusterws}) with $m=n(G_i')$.  
Since the variable $s$ only appears in Eq. (\ref{clusterws}) in the 
form 
\beq
\prod_{i=1}^{k(G')}\Big (q-s+s w^{n(G'_i)} \Big ) =
\prod_{i=1}^{k(G')}\Big (q+s(w-1)\sum_{r=0}^{n(G'_i)-1}w^r \Big ) \ ,
\label{auxeqs}
\eeq
it follows that $Z(G,q,s,v,w)$ and $Ph(G,q,s,w)$ can equivalently be written as
polynomials in the variables $q$, $v$, $w$, and  
\beq
t=s(w-1) \ .
\label{tvar}
\eeq
The advantage of doing this is that it shortens expressions for these
polynomials; however, it renders the symmetries (\ref{zsym}) and (\ref{phsym})
below not manifest in the resultant expressions.

Having shown the connection of $Ph(G,q,s,w)$ to $Z(G,q,s,v,w)$, we observe that
various properties of $Ph(G,q,s,w)$ can be expressed more generally as
corresponding properties of $Z(G,q,s,v,w)$.  From Eq. (\ref{clusterws}) it
follows that the Potts model partition function $Z(G,q,s,v,w)$ satisfies a
basic symmetry relating the values $s$ and $q-s$:
\beq
Z(G,q,s,v,w) = w^n \, Z(G,q,q-s,v,w^{-1})
\label{zsym}
\eeq
so that, in particular, setting $v=-1$, 
\beq
Ph(G,q,s,w) = w^n \, Ph(G,q,q-s,w^{-1}) \ . 
\label{phsym}
\eeq
The symmetry relation (\ref{zsym}) is obvious from a statistical mechanics
context as well as from the formula Eq. (\ref{clusterws}); it is a statement of
the fact that the presence of the magnetic field disfavors or favors the set of
spin values $\sigma_i \in I_s$ relative to the orthogonal complement of spin
values $\sigma_i \in I_s^\perp$, but, up to the prefactor, this is equivalent
to replacing $s$ by $q-s$ and reversing the sign of $H$, i.e., replacing $w$ by
$1/w$.

If the magnetic field is zero, i.e., $w=1$, or $s=0$, so that no spin values
are weighted differently by this field, we have
\beq
Z(G,q,s,v,1) = Z(G,q,0,v,w) = Z(G,q,v) \ , 
\label{zw1andzs0}
\eeq
so that 
\beq
\frac{\partial Z(G,q,s,v,1)}{\partial s} = 0 
\label{dzw1ds}
\eeq
and
\beq
\frac{\partial Z(G,q,0,v,w)}{\partial w} = 0 \ . 
\label{dzz1ds}
\eeq
If $s=q$, so that all spin values receive the same weighting, then 
\beq
Z(G,q,q,v,w) = w^n \, Z(G,q,v) \ . 
\label{zsq}
\eeq
Note that this result also follows by applying the symmetry relation
(\ref{zsym}), so that $Z(G,q,q,v,w)=w^nZ(G,q,0,v,w^{-1})=w^n Z(G,q,v)$. 
Moreover, if the disfavoring is total, i.e., $w=0$, then 
\beq
Z(G,q,s,v,0) = Z(G,q-s,v) \ . 
\label{zw0}
\eeq

   From the definition of $Ph(G,q,s,w)$ as a sum of terms $w^{n_s}$
corresponding to proper $q$-colorings of the vertices of the graph $G$ such
that $n_s$ vertices are assigned colors in the weighted set $I_s$, we can infer
a general inequality.  Let us denote the total set of proper $q$-colorings of
the vertices of $G$ as $\{\sigma\}$ and a subset as $\{\sigma\}_{subset}$, and
let us define $Ph(G,q,s,w)_{subset}$ as the sum of terms $w^{n_s}$ resulting
from the contributions of the subset $\{\sigma\}_{subset}$ of proper
$q$-colorings of the vertices of $G$.  Then, since for the coloring problem at
hand, where $w \ge 0$, each such proper $q$-coloring contributes a non-negative
term to $Ph(G,q,s,w)$, we have the general inequality
\beq
Ph(G,q,s,w) \ge Ph(G,q,s,w)_{subset}  \ . 
\label{phineq}
\eeq

\subsection{Some Properties Connected with Characteristics of Graphs} 

If $G$ has any loop, defined as an edge that connects a vertex to itself, then
a proper $q$-coloring is impossible.  This is because such a $q$-coloring
requires that any two adjacent vertices have different colors, but since the
vertices connected by an edge are adjacent, the presence of a loop in $G$ means
that a vertex is adjacent to itself.  Thus, $Ph(G,q,s,w) = 0$ if $G$ contains a
loop.  Hence, with no loss of generality, in our discussions of $Ph(G,q,s,w)$
we shall restrict our analysis in this paper to loopless graphs $G$.  Thus, in
the text below, where $G=(V,E)$ is characterized as having a non-empty edge set
$E \ne \emptyset$, it is understood that $E$ does not contain any loops.

Another basic property of a chromatic polynomial is that as long as two
vertices are joined by an edge, adding more edges connecting these vertices
does not change the chromatic polynomial. This is clear from the fact that the
chromatic polynomial counts the number of proper $q$-colorings of the vertices
of $G$, and the relevant condition - that two adjacent vertices must have
different colors - is the same regardless of whether one or more than one edges
join these vertices.  Let us define an operation of ``\underline{r}eduction of
multiple \underline{e}dge(s)'' in $G$, denoted $R_E(G)$, as follows: if two
vertices are joined by a multiple edge, then delete all but one of these edges,
and carry out this reduction on all edges, so that the resultant graph $R_E(G)$
has only single edges. Then if $G$ is a graph that contains one or more
multiple edges joining some set(s) of vertices, $P(G,q) = P(R_E(G),q)$. 
Since the same proper $q$-coloring condition holds for the weighted-set
chromatic polynomial, we have
\beq
Ph(G,q,s,w) = Ph(R_E(G),q,s,w) \ . 
\label{phgme}
\eeq

Moreover, if $G$ consists of two disjoint parts, $G_1$ and $G_2$, then
$Z(G,q,s,v,w)$ is simply the product
$Z(G,q,s,v,w)=Z(G_1,q,s,v,w)Z(G_2,q,s,v,w)$, and the same factorization
property holds for the special case $v=-1$ that yields $Ph(G,q,s,w)$.  Hence,
without loss of generality, unless otherwise indicated, we shall restrict our 
discussion here to connected graphs $G$.

\subsection{Properties of Coefficients in Polynomial Expansions}

Next, we prove some general structural properties of $Z(G,q,s,v,w)$ and
$Ph(G,q,s,w)$ that hold for an abitrary graph $G$. From Eqs. (\ref{clusterws})
and (\ref{phclusterws}) one can derive certain factorization properties of
these polynomials. It is convenient to define the notation
\beq
\tilde q = q-s \ . 
\label{tilde}
\eeq
From Eq. (\ref{clusterws}), it follows that we can write $Z(G,q,s,v,w)$ in
several equivalent ways:
\beqs
& & Z(G,q,s,v,w) = \sum_{i,j,\ell=0}^{n} \, \sum_{k=0}^{e(G)} \ 
a_{i,j,k,\ell} \, q^i s^j v^k w^\ell  \ = \
\sum_{i,j,\ell=0}^{n} \, \sum_{k=0}^{e(G)} \ b_{i,j,k,\ell} \, 
q^i s^j y^k w^\ell  \cr\cr
& = &
\sum_{i,j,\ell=0}^{n} \, \sum_{k=0}^{e(G)} \ c_{i,j,k,\ell} \,
\tilde q^{\, i} s^j v^k w^\ell \ = \
\sum_{i,j,\ell=0}^{n} \, \sum_{k=0}^{e(G)} \ d_{i,j,k,\ell} \, 
q^i t^j v^k w^\ell
 \ ,
\label{zgenformwv}
\eeqs
where $a_{i,j,k,\ell}$, $b_{i,j,k,\ell}$, $c_{i,j,k,\ell}$, and
$d_{i,j,k,\ell}$ are integers (and $i, \ j, \ k, \ \ell$ are dummy summation
variables here), Some $a_{i,j,k,\ell}$ and $b_{i,j,k,\ell}$ can be negative,
but the nonzero $c_{i,j,k,\ell}$ and $d_{i,j,k,\ell}$ are positive, as follows
from Eq. (\ref{clusterws}) and Eq. (\ref{auxeqs}) . From these equations, one
infers corresponding ones for $Ph(G,q,s,w)$ by setting $v=-1$, i.e., $y=0$.

For our analysis below and for comparisons with chromatic polynomials, three
types of polynomial expansions will be useful.  Because of the basic symmetry 
(\ref{zsym}), the most useful expansion of $Z(G,q,s,v,w)$ is as a sum of 
powers of $w$ with coefficients, denoted as $\beta_{G,j}(q,s,v)$, 
which are polynomials in $q$, $s$, and $v$: 
\beq
Z(G,q,s,v,w) = \sum_{j=0}^n \beta_{Z,G,j}(q,s,v) \, w^j \ . 
\label{zsumw}
\eeq
The symmetry (\ref{zsym}) implies the following relation among the
coefficients: 
\beq
\beta_{Z,G,j}(q,s,v) = \beta_{Z,G,n-j}(q,q-s,v) \quad 
{\rm for} \ 0 \le j \le n \ . 
\label{zbetasym}
\eeq
In particular, for the special case $v=-1$ of primary interest here, we write 
\beq
Ph(G,q,s,w) = \sum_{j=0}^n \beta_{G,j}(q,s) \, w^j \ , 
\label{phsumw}
\eeq
where
\beq
\beta_{G,j}(q,s) \equiv \beta_{Z,G,j}(q,s,-1)
\label{betabeta}
\eeq
From (\ref{zbetasym}), we have 
\beq
\beta_{G,j}(q,s) = \beta_{G,n-j}(q,q-s) \quad {\rm for} \ 0 \le j \le n \ . 
\label{phbetasym}
\eeq
If $n$ is even, say $n=2m$, then
the middle coefficient is transformed into itself, giving rise to the results
that
\beqs
{\rm If} \ n=2m \ {\rm is \ even, \ then} & \quad & 
\beta_{Z,G,m}(q,s,v) = \beta_{Z,G,m}(q,q-s,v) \cr\cr 
& \quad & \beta_{G,m}(q,s) = \beta_{G,m}(q,q-s)
\label{betamiddle}
\eeqs

From Eq. (\ref{clusterws}), it is clear that the term of highest degree in $w$
arises from products of the $sw^{n(G_i')}$ factors over the various connected
components $G_i'$ for each spanning subgraph $G' \subseteq G$, and then over
the spanning subgraphs $G'$.  This product does not involve $q$, so that 
\beq
\beta_{Z,G,n}(q,s,v) \ {\rm and} \ 
\beta_{G,n}(q,s) \ {\rm are \ independent \ of} \ q \ . 
\label{zbetajns}
\eeq
The $\beta_{Z,G,j}(q,s,v)$ coefficients have especially simple factorization
properties, which we analyze next.  Evaluating Eq. (\ref{zsumw}) at $w=0$,
where only the $w^0$ term remains, and combining this evaluation with the
relation (\ref{zw0}), we derive the result
\beq
\beta_{Z,G,0}(q,s,v)=Z(G,q-s,v) \ . 
\label{zbetaj0}
\eeq

Combining the relation for $w=1$ in Eq. (\ref{zw1andzs0}) with Eq. 
(\ref{zsumw}), we derive a formula for the sum of the coefficients
$\beta_{Z,G,j}(q,s,v)$:
\beq
\sum_{j=0}^n \beta_{Z,G,j}(q,s,v) = Z(G,q,v) \ . 
\label{zw1betasum}
\eeq
Because this sum is independent of $s$, Eq. (\ref{zw1betasum}) 
also yields the differential equation
\beq
\frac{\partial}{\partial s} \, \sum_{j=0}^n \beta_{Z,G,j}(q,s,v) = 0 \ . 
\label{dzbetasumds}
\eeq

Next, we set $s=q$ in Eq. (\ref{zsumw}) and use Eq. (\ref{zsq}).  Since the
resulting expression must be proportional to $w^n$, all of the coefficients of
the terms in $Z(G,q,s,v,w)$ of lower degree in $w$ than $n$ must vanish.
Because these coefficients $\beta_{Z,G,j}(q,s,v)$ are polynomials in $q$ and
$s$ (as well as $v$), this means that they must contain the factor $(q-s)$:
\beq
\beta_{Z,G,j}(q,s,v) \ {\rm and} \ \beta_{G,j}(q,s) \ {\rm contain \ the \ 
factor} \ (q-s) \ {\rm for} \ 0 \le j \le n-1 \ . 
\label{betalowerfactor}
\eeq
Furthermore, using Eq. (\ref{zsq}) for this $s=q$ evaluation, we infer that
$\beta_{Z,G,n}(q,q,v) = Z(G,q,v)$.  However, since, by Eq. (\ref{zbetajns}),
$\beta_{Z,G,n}(q,s,v)$ is independent of $q$ and is only a function of $s$ and
$v$, this implies that
\beq
\beta_{Z,G,n}(q,s,v) = Z(G,s,v)
\label{zbetajn}
\eeq
(so we could drop the argument $q$, but for uniformity with other coefficients
$\beta_{Z,G,j}(q,s,v)$, we shall retain it). 

Now setting $s=0$ reduces $Z(G,q,s,v,w)$ to $Z(G,q,v)$
(cf. Eq. (\ref{zw1andzs0})). Since the $w^0$ term (given in
Eq. (\ref{zbetaj0})) is, by itself, equal to $Z(G,q,v)$ for $s=0$, this means
that all of the other terms proportional to nonzero powers $w^j$, $j=1,...,n$
in $Z(G,q,s,v,w)$ must vanish when $s=0$.  This proves that
\beq
{\rm For} \ 1 \le j \le n, \ \beta_{Z,G,j}(q,s,v) \ {\rm and} \ 
\beta_{G,j}(q,s) \ {\rm contain \ a \  factor \ of } \ s \ . 
\label{zbetaj1n}
\eeq

Various special cases of these results for the weighted-set chromatic
polynomial are obtained by setting $v=-1$ in the requisite equations.  Thus,
Eq. (\ref{zbetaj0}) implies
\beq
\beta_{G,0}(q,s)=P(G,q-s) \ , 
\label{betaj0}
\eeq
and Eq. (\ref{zbetajn}) implies 
\beq
\beta_{G,n}(q,s) = P(G,s) \ . 
\label{betajn}
\eeq

We now focus on $Ph(G,q,s,w)$.  The chromatic number of $G$, denoted $\chi(G)$,
is the minimal number of colors for which one can carry out a proper
$q$-coloring of the vertices of $G$.  Since the proper $q$-coloring constraint
cannot be satisfied for integer $q$ in the interval $0 \le q \le \chi(G)-1$,
the chromatic polynomial $P(G,q)$ vanishes for these values and hence contains
$\prod_{j=0}^{\chi(G)-1}(q-j)$ as a factor. Applying this to Eq. (\ref{betajn})
shows that
\beq
\beta_{G,n}(q,s)\ {\rm contains \ the \ factor} \ \prod_{j=0}^{\chi(G)-1}(s-j) 
\ , 
\label{betanprod}
\eeq
and applying it to Eq. (\ref{betaj0}), taking into account the shift $s \to
q-s$, shows that 
\beq
\beta_{G,0}(q,s) \ {\rm contains \ the \ factor} \ 
\prod_{j=0}^{\chi(G)-1}(q-s-j) \ . 
\label{betaj0factors}
\eeq
In particular, provided that $G=(V,E)$ contains at least one edge, so that
$\chi(G) \ge 2$, we have the results 
\beq
{\rm If} \ E \ne \emptyset, \ {\rm then} \ \beta_{G,n}(q,s) \
{\rm contains \ a \ factor} \ s(s-1)
\label{betajn_factors}
\eeq
and
\beq
{\rm If} \ E \ne \emptyset, \ {\rm then} \ \beta_{G,0}(q,s) \
{\rm contains \ a \ factor} \ (q-s)(q-s-1) \ . 
\label{betaj0_factors}
\eeq
Thus, although the maximal degree of $Ph(G,q,s,w)$ in $w$ is, in general, $n$,
it is less than $n$ if $s=0$ or $s=1$.  In the $s=0$ case, all dependence on
$w$ disappears (cf. Eq. (\ref{phw1andphs0})), while for $s=1$ we have
previously analyzed the maximal degree of $Ph(G,q,1,w)$ for various families of
graphs in Ref. \cite{ph}.  For $s=1$, provided that $G$ contains at least one
edge, $Ph(G,q,w)$ has a factor of $(q-1)$ \cite{ph}.  This is not, in general,
true for $s$ in the interval $2 \le s \le q-1$, and this is one of the ways
that the properties of $Ph(G,q,s,w)$ for $2 \le s \le q-1$ differ from those
for $s=1$.

One can also express $Z(G,q,s,v,w)$ as a polynomial in $q$ with coefficients,
denoted as $\alpha_{Z,G,\ell}(s,v,w)$, which are polynomials in $s$, $v$, and 
$w$:
\beq
Z(G,q,s,v,w) = \sum_{j=0}^n \alpha_{Z,G,n-j}(s,v,w) \, q^{n-j} \ . 
\label{zsumq}
\eeq
Accordingly, with the notation
\beq
\alpha_{G,n-j}(s,w) \equiv \alpha_{Z,G,n-j}(s,-1,w) \ , 
\label{alfalf}
\eeq
we write
\beq
Ph(G,q,s,w) = \sum_{j=0}^n \alpha_{G,n-j}(s,w) \, q^{n-j} \ . 
\label{phsumq}
\eeq
From our discussion above, we have the following results for these latter
coefficients for $Ph(G,q,s,w)$:
\beq
\alpha_{G,n}(s,w) = 1 
\label{alphaqn}
\eeq
and, using also Eq. (\ref{phgme}), 
\beq
\alpha_{G,n-1}(s,w) = ns(w-1) - e(R_E(G)) \ . 
\label{alphaqnminus1}
\eeq
Finally, since the variable $s$ only enters $Z$ via the combination $t=s(w-1)$,
it is also useful to express the coefficients in Eq. (\ref{zsumq}) as
polynomials in $t$, $v$, and $w$, and the coefficients in Eq. (\ref{phsumq}) as
polynomials in $t$ and $w$.  For a given graph $G$, we find that this
usually simplifies the expressions.

A chromatic polynomial $P(G,q)$, written in the form 
\beq
P(G,q) = \sum_{j=0}^{n-k(G)} \alpha_{G,n-j}\, q^{n-j} \ , 
\label{pform}
\eeq
has the property that the signs of the coefficients $\alpha_{G,n-j}$ alternate:
\beq
{\rm sgn}(\alpha_{G,n-j}) = (-1)^j \ , \quad 0 \le j \le n-k(G) \ . 
\label{aj}
\eeq
(where, as before, $k(G)$ denotes the number of components of $G$, and we shall
continue, without loss of generality, to focus on connected graphs, so that
$k(G)=1$.)  This sign alternation property can be proved by iterated
application of the deletion-contraction relation.  Since the weighted-set
chromatic polynomial $Ph(G,q,s,w)$ does not, in general, obey a
deletion-contraction relation, except for the values $w=1$, $w=0$, and $s=0$
for which it reduces to a chromatic polynomial, one does not expect the
corresponding coefficients $\alpha_{G,n-j}(s,w)$ in Eq. (\ref{phsumq}) to have
this sign-alternation property in general, and they do not.  However, we have
proved that if $w$ is in the DFSCP interval $0 \le w < 1$, then the sign
alternation property again holds, i.e.,
\beq
{\rm sgn}(\alpha_{G,n-j}(s,w)) = (-1)^j \quad {\rm for} \ 0 \le w < 1 \ 
{\rm and} \ 0 \le j \le n-1
\label{sa}
\eeq
The technical mathematical details of our proof will be given elsewhere.  For
the borderline cases $w=1$ and $w=0$, as well as for $s=0$ and $s=q$,
$Ph(G,q,s,w)$ reduces to a chromatic polynomial, so the sign-alternation
property is already established. For $j=n$, namely for the $q^0$ term in
$Ph(G,q,s,w)$, the sign alternation also holds for $0 \le w < 1$; here the
coefficient $\alpha_{G,0}(s,w)$ contains the factor $t=s(w-1)$ and hence
vanishes at $w=1$ and $s=0$.

Setting $q=0$ in Eq. (\ref{clusterws}), and recalling the factorization in
Eq. (\ref{auxeqs}), we deduce that 
\beq
Z(G,0,v,s,w) = \alpha_{Z,G,0}(s,v,w) \ 
{\rm contains \ a \ factor \ of} \ t=s(w-1) \ . 
\label{zq0}
\eeq
The same holds, {\it a fortiori}, for the $v=-1$ special case, $Ph(G,0,s,w)$,
i.e., $\alpha_{G,0}(s,w)$ contains a factor of $t=s(w-1)$.

From a study of chromatic polynomials, R. Read observed that the magnitudes of
the coefficients of successive powers of $q^{n-j}$, $0 \le j \le n-k(G)$ in a
chromatic polynomial satisfy a unimodal property \cite{rrev}.  That is, the
magnitudes of these coefficients get successively larger and larger, and then
smaller and smaller, as $j$ increases from 0 to $n-k(G)$.  There is thus a
unique maximal-magnitude coefficient, or two successive coefficients whose
magnitudes are equal. From our calculations of weighted chromatic polynomials
for a number of families of graphs, we have observed that in the interval $0
\le w \le 1$ this property continues to hold.  We therefore state the following
conjecture: {\it Conject.} Let $Ph(G,q,s,w)$ be written as in
Eq. (\ref{phsumq}). Then for real $w$ in the interval $0 \le w \le 1$, the
quantities $(-1)^j\alpha_{G,n-j}(s,w)$, $0 \le j \le n$, are positive and
satisfy the unimodal property, i.e., $(-1)^j\alpha_{G,n-j}(s,w)$ get
progressively larger and larger, and a maximal value is reached for a given
$j$, or for two successive $j$ values, and then the quantities
$(-1)^j\alpha_{G,n-j}(s,w)$ get progressively smaller, as $j$ increases from 0
to $n$.

\subsection{Measure of Deviation from Deletion-Contraction Relation}

For a graph $G$, let us denote the graph obtained by deleting an edge $e \in E$
as $G-e$ and the graph obtained by deleting this edge and identifying the two
vertices that had been connected by it as $G/e$.  The Potts model partition
function satisfies the deletion-contraction relation (DCR) 
\beq
Z(G,q,v) = Z(G-e,q,v)+vZ(G/e,q,v)
\label{zdelcon}
\eeq
and, setting $v=-1$, the chromatic polynomial thus satisfies the DCR 
\beq
P(G,q,v) = P(G-e,q)-P(G/e,q) \ . 
\label{pdelcon}
\eeq
However, in general, neither $Z(G,q,s,v,w)$ nor $Ph(G,q,s,w)$ satisfies the
respective deletion-contraction relation.  For the special cases $w=1$ and
$s=0$ for which $Z(G,q,s,v,w)$ and $Ph(G,q,s,w)$ reduce to $Z(G,q,v)$ and
$P(G,q)$, and for the special case $w=0$ for which $Z(G,q,s,v,w)$ and
$Ph(G,q,s,w)$ reduce to $Z(G,q-s,v)$ and $Ph(G,q-s)$, respectively, they do
satisfy deletion-contraction relation.  Hence, the deviations from such a
relation for $Z(G,q,s,v,w)$ and $Ph(G,q,s,w)$ vanish in these three cases.
For a given $G$, it is of interest to examine the quantities that measure the
deviation from the DCR, namely
\beq
[\Delta Z(G,e,q,s,v,w)]_{DCR}=Z(G,q,s,v,w) 
-\Big [ Z(G-e,q,s,v,w)+vZ(G/e,q,s,v,w)\Big ] 
\label{zdelcondif}
\eeq
and
\beq
[\Delta Ph(G,e,q,s,w)]_{DCR} \equiv [\Delta Z(G,e,q,s,-1,w)]_{DCR} \ . 
\label{phdelcondif}
\eeq
From our discussion above, it follows that $[\Delta Z(G,e,q,s,v,w)]_{DCR}=0$,
and hence also $\Delta Ph(G,e,q,s,w)]_{DCR}=0$, for $w=1$, $w=0$, and $s=0$;
therefore, since these functions are polynomials in these variables, they
contain a factor $sw(w-1)$.  Moreover, since the condition $v=0$ is equivalent
to the absence of any edges, whence $Z=(q+t)^n$ and the deletion-contraction
relation is satisfied trivially, $[\Delta Z(G,e,q,s,v,w)]_{DCR}$ also
vanishes for $v=0$.  Thus, in general,
\beq
[\Delta Z(G,e,q,s,v,w)]_{DCR} \ {\rm contains \ the \ factor} \ svw(w-1) \ , 
\label{deltafactors}
\eeq
and $[\Delta Ph(G,e,q,s,w)]_{DCR}$ contains a factor of $sw(w-1)$. As an
illustration, using our explicit calculations for $n$-vertex line graphs $L_n$
and circuit graphs $C_n$, we find the following results. For the first two
graphs, $L_3$ and $C_3$, the deletion and contraction on any edge gives the
same result, so we need not specify which edge is involved.  We find, for any
edge $e$,
\beq
[\Delta Z(L_2,e,q,s,v,w)]_{DCR} = svw(w-1) \ , 
\label{zline2_delcondif}
\eeq
\beq
[\Delta Z(L_3,e,q,s,w)]_{DCR} = svw(w-1)\Big [ s(w-1)+wv+q \Big ]  \ , 
\label{zline3_delcondif}
\eeq
and
\beq
[\Delta Z(C_3,e,q,s,v,w)]_{DCR} = svw(w-1)\Big [ wv^2+2wv+s(w-1)+q \Big ] \ . 
\label{phcyc4_delcondif}
\eeq
As before, the corresponding $[\Delta Ph(G,e,q,s,w)]_{DCR}$ expressions are
obtained by setting $v=-1$ in these equations.  It is straightforward to
calculate similar differences $[\Delta Z(G,e,q,s,v,w)]_{DCR}$ for graphs with
more vertices and edges, but these are sufficient for our illustration.

\subsection{Distinguishing Between Various Equivalence Classes of Graphs}

An important property of the weighted-set chromatic polynomial $Ph(G,q,s,w)$ is
the fact that it can distinguish between certain graphs that yield the same
chromatic polynomial $P(G,q)$.  This is true for all $w$ and $s$ values except
the special values $w=1$, $w=0$, $s=0$, and $s=q$, for which $Ph(G,q,s,w)$ is
reducible to a chromatic polynomial.  More generally, an important property of
the partition function of the Potts model in a set of nonzero external magnetic
fields of the form (\ref{hp}), $Z(G,q,s,v,w)$, is that this function can
distinguish between graphs that yield the same zero-field Potts model partition
function, $Z(G,q,s,v,1)=Z(G,q,v)$.  Two graphs $G$ and $H$ are defined as (i)
Tutte-equivalent if they have the same Tutte polynomial, or equivalently, the
same zero-field Potts model partition function, $Z(G,q,v)$, and (ii)
chromatically equivalent if they have the same chromatic polynomial, $P(G,q)$.
Here we recall that the Tutte polynomial $T(G,x,y)$ of a graph $G$ is defined
as
\beq
T(G,x,y) = \sum_{G' \subseteq G} (x-1)^{k(G')-k(G)}(y-1)^{c(G')}
\label{t}
\eeq
where $G'$ is a spanning subgraph of $G$ and $c(G')$ denotes the number of
(linearly independent) cycles in $G'$.  This polynomial is equivalent to the 
zero-field Potts model partition function, via the relation
\beq
Z(G,q,v) = (x-1)^{k(G)}(y-1)^n T(G,x,y) \ . 
\label{ztrel}
\eeq
where $y=v+1$ as in Eq. (\ref{kdef}) and 
\beq
x=1+\frac{q}{v} \ . 
\label{x}
\eeq

We give some examples.  Recall the definition that a tree graph is a connected
graph that contains no circuits (cycles).  The set of tree graphs with $n$
vertices, generically denoted $\{T_n\}$, forms a Tutte equivalence class, with
$T(T_n,x,y)=x^{n-1}$, or equivalently,
\beq
Z(T_n,q,v)=q(q+v)^{n-1} \ . 
\label{ztree}
\eeq
However, $Z(G,q,s,v,w)$ is able to distinguish between different tree graphs in
a Tutte-equivalence class.  Because $Z(G,q,s,v,w)$ reduces to a zero-field
Potts model partition function for $w=1$, $w=0$, $s=0$, and $s=q$, it follows
that the difference between $Z(G,q,s,v,w)$ and $Z(H,q,s,v,w)$ for two
Tutte-equivalent graph $G$ and $H$ must vanish if $w=1$, $w=0$, $s=0$, or
$s=q$. This difference also vanishes for $v=0$, because in this case the only
spanning subgraph that contributes to the sum in Eq. (\ref{clusterws}) is the
one with no edges, which is the same for any (connected) Tutte-equivalent $G$
and $H$.  Since these are all polynomials, it thus follows that
\beq
Z(G,q,s,v,w)-Z(H,q,s,v,w) \ {\rm contains \ the \ factor} \ s(q-s)vw(w-1) \ . 
\label{phgfdiff}
\eeq
As an illustration, using our results for $Z(L_4,q,s,v,w)$ 
and $Z(S_4,q,s,v,w)$, we have 
\beq
Z(S_4,q,s,v,w)-Z(L_4,q,s,v,w)=s(q-s)v^2w(w-1)^2
\label{zstar4_minus_zline4}
\eeq
and consequently
\beq
Ph(S_4,q,s,w)-Ph(L_4,q,s,w)=s(q-s)w(w-1)^2 \ . 
\label{phstar4_minus_phline4}
\eeq

The zero-field Potts model partition function $Z(G,q,v)$, or equivalently, 
the Tutte polynomial $T(G,x,y)$, encodes information on the number of 
(linearly independent) cycles contained in the graph $G$, as is evident from
the definition (\ref{t}).  Define two scaled variables as 
\beq
q' \equiv \frac{q}{s} \ , \quad v' \equiv \frac{v}{s} \ . 
\label{qvprime}
\eeq
Let us us consider a graph, denoted $G_{nc}$, which contains no cycles (nc),
i.e., which has $c(G)=0$.  A connected graph of this type is a tree graph,
while a general graph is called a forest. For a graph $G_{nc}$ we find the
following scaling relation:
\beq
Z(G_{nc},q,s,v,w) = s^n Z(G_{nc},q',1,v',w) \ . 
\label{zscaled}
\eeq
This is proved as follows.  We start with the cluster formula (\ref{clusterws})
and rewrite this as 
\beq
Z(G_{nc},q,s,v,w) = \sum_{G' \subseteq G_{nc}} 
(v')^{e(G')} \, s^{e(G')+k(G')} \, 
\prod_{i=1}^{k(G')} \Big ( q'-1 + w^{n(G'_i)} \Big ) \ . 
\label{clusterwscaled}
\eeq
We next use the relation, which holds for any graph $G'$, 
\beq
c(G') + n(G') = e(G') + k(G') 
\label{crel}
\eeq
and the fact that $n(G')=n(G) \equiv n$ to rewrite the factor 
$s^{e(G')+k(G')}$ as $s^{c(G')+n}$.  Since $G_{nc}$ has no cycles, it follows
that the same is true for any subgraph of $G_{nc}$, in particular, the spanning
subgraph $G'$, so $c(G')=0$.  Hence, we can move the factor of $s^n$ in
front of the summation, and we have 
\beqs
Z(G_{nc},q,s,v,w) & = & s^n \sum_{G' \subseteq G_{nc}} 
(v')^{e(G')} \, \prod_{i=1}^{k(G')} \Big ( q'-1 + w^{n(G'_i)} \Big ) \cr\cr
                  & = & s^n \, Z(G_{nc},q',1,v',w) \ . 
\label{clusterwscaled2}
\eeqs
$\Box$  

Hence, the difference 
\beq
[\Delta Z(G,q,s,v,w)]_{cycles} = Z(G,q,s,v,w) - s^n Z(G,q',1,v',w)
\label{deltaz_cycles}
\eeq
provides a measure of the number of cycles in $G$.  For example, using our
general result for $Z(C_n,q,s,v,w)$ given below in Eq. (\ref{zcn}), we
calculate 
\beq
[\Delta Z(C_n,q,s,v,w)]_{cycles} = \frac{(s-1)(q-s+sw^n)v^n}{s}
\label{deltazcncycles}
\eeq

Moreover, the proper $q$-coloring condition implies that if two different
graphs $G$ and $H$ differ only in having different numbers of edges connecting
a pair of adjacent vertices, for one or more such pairs, so that
$R_E(G)=R_E(H)$, then $Ph(G,q,s,w)=Ph(H,q,s,w)$.  A simple example is provided
by the line and circuit graphs with $n=2$ vertices, $L_2$ and $C_2$, the latter
of which has a double edge connecting the two vertices. Using our results in
Eqs. (\ref{zline2}) and (\ref{zc2}), we calculate the difference
\beq
Z(C_2,q,s,v,w)-Z(L_2,q,s,v,w)=v(v+1)\left [ q+s(s-1)(w+1) \right ] \ . 
\label{zc2_minus_zline2}
\eeq
The fact that the difference in Eq. (\ref{zc2_minus_zline2}) vanishes for
$v=-1$, i.e., that $Ph(L_2,q,s,w)=Ph(C_2,q,s,w)$, is a special case of the
general result (\ref{phgme}). 

In the context of graph coloring, since $s \in I_s$, if one sets $q$ to a
particular value, this implicitly sets a corresponding upper bound on $s$. In
particular, if $q=1$, then $s$ can take on only the values 0 or 1, and these
are related by the symmetry (\ref{zsym}).  For $s=0$, we have, by Eq.
(\ref{zw1andzs0}), that
\beq
Z(G,1,0,v,w)=Z(G,1,v)=y^{e(G)}w^n \ , 
\label{zs0}
\eeq
where $y=v+1$. For $s=1$, applying Eq. (\ref{zsym}), we have
\beq
Z(G,1,1,v,w)=w^nZ(G,1,0,v,w^{-1})=w^nZ(G,1,v)=y^{e(G)}w^{2n} \ .
\label{zs1}
\eeq
If $G$ has at least one edge, then the right-hand sides of both Eq. (\ref{zs0})
and Eq. (\ref{zs1}) vanish for the case $y=0$ ($v=-1$) that yields the
weighted-set chromatic polynomal. In order for two graphs $G$ and $H$ to be
chromatically equivalent, a necessary condition is that they must have the same
number of vertices, $n(G)=n(H)$, since the degree in $q$ of $P(G,q)$ is
$n(G)$. An elementary property of the chromatic polynomial $P(G,q)$, proved by
iterative application of the deletion-contraction theorem, is that the
coefficient of the $q^{n(G)-1}$ term is $-e(R_E(G))$.  Therefore, another
necessary condition that two graphs $G$ and $H$ must satisfy in order to be
chromatically equivalent is that $e(R_E(G))=e(R_E(H))$.  If $G$ contains at
least one edge, then $Ph(G,1,s,w)=0$.  Note here that since $s$ is bounded
above by $q$, it follows that if $q=1$, then $s$ can only take on the values
$s=0$ or $s=1$.  Hence, if $G$ and $H$ are chromatically equivalent, then
either (i) neither contains any edges, in which case
$Ph(G,q,s,w)=Ph(H,q,s,w)=(q+t)^n$, where $n=n(G)=n(H)$, or (ii) if $G$, and
hence $H$, contains at least one edge, then $Ph(G,1,s,w)=Ph(H,1,s,w)=0$.
Hence, if $G$ and $H$ are chromatically equivalent and contain at least one
edge, then the difference $Ph(G,q,s,w)-Ph(H,q,s,w)$ contains a factor that
vanishes when $q=1$.  Because of the implicit condition on $s$ for a given $q$,
this factor is not, in general, $(q-1)$.  As an example, the difference
$Ph(S_4,q,s,w)-Ph(L_4,q,s,w)$ in Eq. (\ref{phstar4_minus_phline4}) contains the
factor $s(q-s)$.  For $q=1$, the values of $s$ are implicitly restricted to
$s=0$ and $s=1$.  For either of these choices, the factor, and hence the
difference, vanishes.

A remark concerning duality for planar graphs is also in order here.  Let
$G=(V,E)$ be a planar graph, and denote its planar dual by $G^*$.  The
chromatic polynomial $P(G,q)$ counts not just the proper $q$-colorings of the
vertices of $G$, but also, and equivalently, the proper $q$ colorings of the
faces of $G^*$.  Similarly, for this planar graph $G$, the weighted-set
chromatic polynomial $Ph(G,q,s,w)$ describes not just the weighted-set proper
$q$-colorings of the vertices of $G$ but also, and equivalently, the
weighted-set proper $q$-colorings of the faces of $G^*$.

\subsection{Lower Bounds on $Ph(G,q,s,w)$} 

We derive some bounds on $Ph(G,q,s,w)$ for certain types of graphs.  Our method
for this will be to calculate the contribution to $Ph(G,q,s,w)$ resulting from
a certain procedure for performing proper $q$-colorings of the graph $G$.  By
the general formula expressing $Ph(G,q,s,w)$ as the $v=-1$ special case of the
Potts model partition function $Z(G,q,s,v,w)$ together with the formulation of
this partition function as a sum over spin (or equivalently, color)
configurations, Eqs. (\ref{z}) with (\ref{ham}), it follows that there are
other color configurations in addition to the particular one that we consider,
contributing (positive terms) to $Ph(G,q,s,w)$.  Therefore each particular
proper $q$-coloring procedure that we consider yields a lower bound on
$Ph(G,q,s,w)$.  The specific proper $q$-coloring procedure that provides a good
lower bound to $Ph(G,q,s,w)$ depends on the type of graph $G$, the values of
$q$, $s$, and $w$.

Let us consider a bipartite graph $G_{bip}$, defined as a graph whose vertex
set $V$ can be partitioned into subsets $V_1$ and $V_2$ such that a vertex in
$V_1$ has edges that connect it only to a vertex or vertices in $V_2$ and vice
versa.  An equivalent condition for a graph to be bipartite is that its
chromatic number $\chi(G_{bip})=2$.  As above, we denote the number of vertices
in $G$, as $n(G) \equiv n$ and, further, the number of vertices in $V_1$ and
$V_2$ as $n_1$ and $n_2$, respectively. With no loss of generality, we label
these subsets of vertices so that $n_1 \le n_2$. These numbers $n_1$ and $n_2$
may be comparable or may be quite different.  For example for a lattice graph
such as the circuit graph, the square, honeycomb, simple cubic, or
body-centered cubic lattices, with periodic boundary conditions that preserve
the bipartite nature of the lattices, $n_1=n_2$.  However, for the star graph
$S_n$, $V_1$ consists of the central vertex, so that $n_1=1$, while $V_2$ is
comprised of all of the vertices on the ends of the edges forming the rays of
the star, so $n_2=n-1$.  For an $S_n$ graph with $n >> 1$, it follows that $n_2
>> n_1$.

For the following, we assume that $q \ge 2$ so that a proper $q$-coloring of
the bipartite graph $G_{bip}$ is possible.  If $w=1$, then a well-known
elementary lower bound on $P(G_{bip},q)$ is obtained by (i) assigning a single
color to all of the vertices in $V_1$ and (ii) independently choosing a color
out of the remaining $q-1$ for each of the vertices in $V_2$.  There are
$q(q-1)^{n_2}$ ways of doing this.  Since, in general, there are also other
color configurations contributing to a proper $q$-coloring of $G_{bip}$, this
yields the lower bound
\beq
P(G_{bip},q) \ge q(q-1)^{n_2} \ . 
\label{pgbound}
\eeq

For the weighted-set chromatic polynomial, the situation is more
complicated.  With no loss of generality, we again label the vertex subsets so
that $n_1 \le n_2$.  We also take $q \ge 2$ so that a proper (weighted) 
$q$-coloring is possible and also assume that $s \ge 2$ (and $s \le
q$, as discussed above).  Then for sufficiently large $w > 1$ in the FSCP
interval, one lower bound on $Ph(G_{bip},q,s,w)$ is obtained by (i) assigning
one color from the favorably weighted set $I_s$ to all of the vertices in
$V_1$, and then (ii) independently, for each vertex in $V_2$, assigning a color
from among the remaining $s-1$ colors in $I_s$.  This combined color assignment
can be made in $s(s-1)^{n_2}$ ways, and yields a contribution $s(s-1)^{n_2}
w^n$ to $Ph(G_{bip},q,s,w)$.  From the inequality (\ref{phineq}), it then
follows that
\beq
Ph(G_{bip},q,s,w) \ge s(s-1)^{n_2} w^n \ . 
\label{phboundlargew}
\eeq
(Since $n_1 \le n_2$, this is an equivalent or better bound than the one
obtained by making the above color assignments with $V_1$ and $V_2$ reversed,
viz., $Ph(G_{bip},q,s,w) \ge s(s-1)^{n_1} w^n$.)  

On the other hand, for $w$ in the DFSCP interval $0 \le w < 1$, it can be
preferable to minimize the number of vertices with colors in $I_s$ in order to
maximize the contribution to $Ph(G_{bip},q,s,w)$.  For sufficiently small
(positive) $w$, provided that $q \ge s+2$, a lower bound on $Ph(G_{bip},q,s,w)$
is then obtained by (i) assigning a single color from $I_s^\perp$ to all of the
vertices of $V_1$, and (ii) independently, for each vertex of $V_2$, assigning
a color from among the remaining $(q-s-1)$ colors in $I_s^\perp$.  This
combined color assignment can be made in $(q-s)(q-s-1)^{n_2}$ ways.  Invoking
the inequality (\ref{phineq}) again, we have
\beq
Ph(G_{bip},q,s,w) \ge (q-s)(q-s-1)^{n_2} \ . 
\label{phboundsmallw}
\eeq
If $w$ is only slightly less than unity, say $1-\epsilon < w < 1$ for
sufficiently small positive $\epsilon$, provided also that $q \ge s+1$, a
different type of lower bound on $Ph(G_{bip},q,s,w)$ can be obtained by the
following proper $q$-coloring procedure: (i) one assigns a single color from
$I_s$ to all of the vertices of $V_1$ and (ii) independently for each vertex in
$V_2$, one assigns a color from $I_s^\perp$.  There are $s(q-s)^{n_2}$ ways of
doing this, and the resultant term in $Ph(G_{bip},q,s,w)$ is
$sw^{n_1}(q-s)^{n_2}$.  Using the inequality (\ref{phineq}) again, one thus
infers the lower bound
\beq
Ph(G_{bip},q,s,w) \ge sw^{n_1}(q-s)^{n_2} \ . 
\label{phboundmoderatew}
\eeq
Which of these lower bounds is the best depends in detail on $G_{bip}$ (in
particular, on $n_1$ and $n_2$), $q$, $s$, and $w$.  It is easy to
generalize these lower bounds to multipartite graphs.

\section{Calculations of $Z(G,q,s,v,w)$ and $Ph(G,q,s,w)$ for Some Families of 
Graphs}

In this section we give some illustrative explicit calculations of
$Z(G,q,s,v,w)$ and $Ph(G,q,s,w)$ for various families of graphs.  Although we
generally consider connected graphs, we note that for the graph $N_n$
consisting of $n$ vertices with no edges,
\beq
Z(N_n,q,s,v,w)=Ph(N_n,q,s,w)=(q+t)^n \ . 
\label{znulln}
\eeq

\subsection{Line Graphs $L_n$}

The line graph (also called path graph) $L_n$ is the graph consisting of $n$
vertices with each vertex connected to the next one by one edge.  In general,
$\alpha_{Z,L_n,n-1}(q,s,v)=nt+(n-1)v$.  We proceed to give some explicit
results for $Z(L_n,q,s,w)$ for various values of $n$. The case $L_1=N_1$ is
already covered by Eq. (\ref{znulln}).  For the first few $n$ values, we also
give the expansions in terms of powers of $q$ and, for this latter expansion,
we use the variables $q$, $t$, and $w$ instead of $q$, $s$, and $w$, because
this makes the expressions shorter:
\beqs
Z(L_2,q,s,v,w) & = & s(s+v)w^2 + 2s(q-s)w +(q-s)(q-s+v) \cr\cr
               & = & q^2+(2t+v)q + t\Big [t+v(w+1) \Big ]
\label{zline2}
\eeqs
\beqs
Z(L_3,q,s,v,w) & = & s(s+v)^2w^3+s(q-s)(3s+2v)w^2 \cr\cr 
               & + & s(q-s)\Big [3(q-s)+2v \Big ]w + (q-s)(q-s+v)^2 \cr\cr
               & = & q^3+(3t+2v)q^2+(3t^2+2vtw+4vt+v^2)q \cr\cr
               & + & t(v^2w^2+2vtw+wv^2+t^2+2vt+v^2) \ . 
\label{zline3}
\eeqs
For $L_4$ we give only the expansion in powers of $w$, since the equivalent 
expansion in powers of $q$ becomes somewhat lengthy: 
\beqs
Z(L_4,q,s,v,w) & = & s(s+v)^3w^4 + 2s(q-s)(s+v)(2s+v)w^3 \cr\cr
           & + & s(q-s)\Big [ -3(s^2+(q-s)^2) + 3q(q+v)+2v^2\Big ]w^2 \cr\cr
           & + & 2s(q-s)(q-s+v)[2(q-s)+v]w + (q-s)(q-s+v)^3 \ . 
\label{zline4}
\eeqs

\subsection{Star Graphs $S_n$}

A star graph $S_n$ consists of one central vertex with degree $n-1$ connected
by edges with $n-1$ outer vertices, each of which has degree 1.  (The context
will always make clear the difference between this symbol for the $n$-vertex
star graph and the symbol $S_n$ for the symmetric group on $n$ objects.)  The
graph $S_2$ is degenerate in the sense that it has no central vertex but
instead coincides with $L_2$.  The graph $S_3$ is nondegenerate, and coincides
with $L_3$, while the $S_n$ for $n \ge 4$ are distinct graphs not coinciding
with those of other families. For $n \ge 2$, the chromatic number is
$\chi(S_n)=2$. By the use of combinatoric coloring methods, we have derived the
following general formula for $Z(S_n,q,s,v,w)$:
\beq
Z(S_n,q,s,v,w)=\sum_{j=0}^{n-1} {n-1 \choose j} \, v^j \, (\tilde q+sw^{j+1}) 
\, (\tilde q+sw)^{n-1-j} \ , 
\label{zstar}
\eeq
where $\tilde q = q-s$, as given in Eq. (\ref{tilde}).  Evaluating
Eq. (\ref{zstar}) for $v=-1$ yields $Ph(S_n,q,s,w)$. As an explicit example,
for the graph $S_4$, we calculate
\beqs
Z(S_4,q,s,v,w) & = & Z(T_4,q,v)w^4+s(q-s)(4s^2+6sv+3v^2)w^3 \cr\cr
               & + & 3s(q-s)[2s(q-s)+qv]w^2
        +s(q-s)\Big [ 4(q-s)^2+6(q-s)v+3v^2 \Big ] w 
\cr\cr
               & + & Z(T_4,q-s,v) 
\label{phstar4}
\eeqs
where $Z(T_n,q,v)$ was given in Eq. (\ref{ztree}).

\subsection{Complete Graphs $K_n$}

The complete graph $K_n$ is the graph with $n$ vertices such that each vertex
is connected to every other vertex by one edge. The chromatic number is 
$\chi(K_n)=n$ and the number of edges is $e(K_n)={n \choose 2}$.  
Let us introduce the compact notation $x_\theta \equiv x \theta(x)$, where
$\theta(x)$ is the step function from ${\mathbb R} \to \{0,1\}$ defined 
as $\theta(x)=1$ if $x > 0$ and $\theta(x) = 0$ if $x \le 0$. 
We have derived the following theorem giving a general formula for 
$Ph(K_n,q,s,w)$:
\beq
Ph(K_n,q,s,w) = \sum_{\ell=0}^n \beta_{K_n,\ell}(q,s) \, w^\ell
\label{phkn}
\eeq
where
\beq
\beta_{K_n,\ell} = {n \choose \ell} \, \Big [ \prod_{j=0}^{(\ell-1)_\theta}
  (s-j) \Big ] \Big [ \prod_{m=0}^{(n-\ell-1)_\theta} (q-s-m) \Big ] \ . 
\label{betaknell}
\eeq
Proof: \quad This result is proved by a combinatoric coloring argument.
Accordingly, we take $q$ to be a non-negative integer.  The resultant
Eqs. (\ref{phkn}) and (\ref{betaknell}) allow the extension of $q$ to
${\mathbb R}$ (and ${\mathbb C}$).  First, if $q < \chi(K_n)=n$, then
$Ph(K_n,q,s,w)$ vanishes identically.  Hence, we shall formally take $q \ge n$
to begin with; once we have obtained the results (\ref{phkn}) and
(\ref{betaknell}), it will be seen that they allow an extension of $q$ away
from this range. If $s \ge n$, then one can assigning $n$ different colors to
the $n$ vertices of $K_n$ from the set $I_s$, and this gives rise to a term
with degree $n$ in $w$.  To determine the coefficient of this term, we
enumerate the number of ways this color assignment can be made.  We pick a
given vertex and assign some color from $I_s$ to this vertex, which we can do
in any of $s$ ways.  Then we go on to the next vertex and assign one of the
remaining $s-1$ colors in $I_s$ to that vertex, and so on for the $n$ vertices.
The number of ways of making this color assignment, i.e., the coefficient of
the term in $Ph(K_n,q,s,w)$ of maximal degree in $w$, viz., $w^n$, is therefore
\beq
\beta_{K_n,n}(q,s)= \prod_{j=0}^{n-1} (s-j) = P(K_n,s) \ . 
\label{betaknn}
\eeq
The fact that this coefficient is $P(K_n,s)$ agrees with the $v=-1$ special
case of the general result of Eq. (\ref{zbetajn}).  Similarly, the term of
order $w^0$ is obtained by assigning $n$ different colors to the $n$ vertices
of $K_n$ from the orthogonal set $S^\perp$.  By reasoning analogous to that
given above, it follows that the number of ways of doing this is given by
replacing $s$ by $q-s$ in Eq. (\ref{betaknn}), so 
\beq
\beta_{K_n,0}(q,s)= \prod_{j=0}^{n-1} (q-s-j) = P(K_n,q-s) \ . 
\label{betakn0}
\eeq
Having illustrated the logic on these two extremal terms, let us next consider
the general $w^\ell$ term with $0 \le \ell \le n$.  This term arises from color
assignments in which we pick $\ell$ different colors from the set $I_s$ and
assign them to $\ell$ of the $n$ vertices of $K_n$, and then $n-\ell$ different
colors from the orthogonal complement set $S^\perp$, which are assigned to the
remaining $n-\ell$ vertices.  The number of ways of doing this is
\beq
\beta_{K_n,\ell} = \Big [ \prod_{j=0}^{(\ell-1)_\theta} (s-j) \Big ] 
\Big [ \prod_{m=0}^{(n-\ell-1)_\theta} (q-s-m) \Big ] \ . 
\label{knassignments}
\eeq
This proves the result in Eqs. (\ref{phkn}) and (\ref{betaknell}). $\Box$ \quad

Evidently, with the polynomial $Ph(K_n,q,s,w)$ as specified in these equations,
one can extend $q$ and $s$ away from non-negative integer values.  Our result
in Eqs. (\ref{phkn}) and (\ref{betaknell}) generalizes the result for the case
$s=1$ given in \cite{ph}.  As is evident, for $w=1$ or $s=0$, $Ph(K_n,q,s,w)$
reduces to the (usual, unweighted) chromatic polynomial
\beq
P(K_n,q) = \prod_{j=0}^{n-1} (q-j) \ . 
\label{pkn}
\eeq

A corollary of Eqs. (\ref{phkn}) and (\ref{betaknell}) is that
\beq
{\rm If} \ s < n \ , \ {\rm then} \ \beta_{K_n,j}(q,s)=0 \ {\rm for} \ s < j
\le n 
\label{betaknjzero}
\eeq
and hence
\beq
{\rm deg}_w(Ph(K_n,q,s,w))= {\rm min}(n,s)  \ . 
\label{degwphkn}
\eeq

Having calculated $Ph(K_n,q,s,w)$, it is appropriate to discuss here another
aspect in which the weighted-set chromatic polynomial differs from the (usual,
unweighted) chromatic polynomial.  Let us consider a graph $G$ that has the
property of being composed of the union of two subgraphs, $G = G_1 \cup G_2$,
such that $G_1 \cap G_2 = K_m$ for some $m$.  In the rest of this paragraph, we
assume that $G$ has this property.  Then $P(G,q)$ satisfies the relation
\beq
P(G,q) = \frac{P(G_1,q)P(G_2,q)}{P(K_m,q)} \ . 
\label{intersectiontheorem}
\eeq
(This is sometimes called the complete-graph intersection theorem (KIT) for
chromatic polynomials.)  In contrast, in general, $Ph(G,q,s,w)$ is not equal to
$Ph(G_1,q,s,w)Ph(G_2,q,s,w)/Ph(K_m,q,s,w)$.  This equality holds only for the
four values $w=1$, $w=0$, $s=0$, and $s=q$ where $Ph(G,q,s,w)$ reduces to a
chromatic polynomial.  As a measure of the deviation from equality, we define
\beq
[\Delta Ph(G,q,s,w)]_{KIT} \equiv Ph(G,q,s,w) - 
\frac{Ph(G_1,q,s,w)Ph(G_2,q,s,w)}{Ph(K_m,q,s,w)} \ . 
\label{kit_diff}
\eeq
The vanishing of $[\Delta Ph(G,q,s,w)]_{KIT}$ for $w=1$, $w=0$, and $s=0$ is
obvious.  To show that this vanishes for $s=q$, we use the relation 
(\ref{zsq}) and obtain 
\beq
[\Delta Ph(G,q,q,w)]_{KIT} = w^n \Big [ P(G,q) - 
 \frac{P(G_1,q)P(G_2,q)}{P(K_m,q)} \Big ] = 0 \ . 
\label{delta_kit_aux}
\eeq
Combining these results with the property that $[\Delta Ph(G,q,s,w)]_{KIT}$ is
a rational function in its arguments, we have thus shown that
\beq
[\Delta Ph(G,q,s,w)]_{KIT} \ {\rm contains \ the \ factor} \ s(q-s)w(w-1) \ . 
\label{kit_diff_factor}
\eeq
We give two illustrations.  The line graph $L_3$ has the property
of being comprised of two $L_2$ graphs intersecting on $L_1 = K_1$.  Using (the
$v=-1$ special cases of) our results in Eqs. (\ref{znulln}), (\ref{zline2}),
and (\ref{zline4}), we calculate
\beq
[\Delta Ph(L_3,q,s,w)]_{KIT} = \frac{s(q-s)w(w-1)^2}{q+s(w-1)} \ . 
\label{phline3_kit_diff}
\eeq
Similarly, the graph $L_4$ can be decomposed into $L_3$ and $L_2$ subgraphs
that intersect on an $L_1=K_1$ graph.  Using our results in Eq. (\ref{znulln})
and (\ref{zline2})-(\ref{zline4}), we calculate
\beq
[\Delta Ph(L_4,q,s,w)]_{KIT} = \frac{s(q-s)w(w-1)^2
\Big [ q+s(w-1)-(w+1) \Big ]}{q+s(w-1)} \ . 
\label{phline4_kit_diff}
\eeq
A slightly more complicated case is the 4-vertex graph $C_{4d}$ consisting of 
a box with one diagonal edge added.  This graph has the structure of two
$C_3=K_3$ subgraphs intersecting on the diagonal edge graph, $L_2=K_2$.  For
this graph we calculate 
\beq
[\Delta Ph(C_{4d},q,s,w)]_{KIT} = 2s(q-s)w(w-1)^2 \Big [ 
1 - \frac{2(q-1)w}{Ph(K_2,q,s,w)} \Big ] \ . 
\label{phc4d_kit_diff}
\eeq
We see that each of these differences $[\Delta Ph(G,q,s,w)]_{KIT}$ satisfies
the general factorization property of Eq. (\ref{kit_diff_factor}).

\section{$Z(G,q,s,v)$ and $Ph(G,q,s,w)$ for Cyclic Strip Graphs}

\subsection{General Structure} 

Refs. \cite{hl,zth} have given a general structural formula for $Z(G_s, L_y
\times m,BC,q,s,v,w)$ on strip graphs $G_s$ of width $L_y$ vertices and
length $L_x$, with cyclic or M\"obius boundary conditions (BC's) for the case
$s=1$. Here we discuss the generalizations to arbitrary (integer) $s$ in the
interval $0 \le s \le q$.  For cyclic strip graphs $G_s$ of this type we have 
\beq
Z(G_s, L_y \times m,cyc.,q,s,v,w) 
 = \sum_{d=0}^{L_y} \tilde c^{(d)} \sum_{j=1}^{n_{Zh}(L_y,d,s)}
[\lambda_{Z,G_s,L_y,d,j}(q,s,v,w)]^m \ , 
\label{zsumcyc}
\eeq
where $m=L_x$ for strips of the square and triangular lattices and $m=L_x/2$
for strips of the honeycomb lattice. The coefficients $\tilde c^{(d)}$ are
given by
\beq
\tilde c^{(d)} \equiv c^{(d)}(\tilde q) = 
\sum_{j=0}^d (-1)^j {2d-j \choose j}\tilde q^{\, d-j} 
\label{ctilde}
\eeq
where $\tilde q = q-s$, as specified in Eq. (\ref{tilde}).  The first few of
these coefficients are $\tilde c^{(0)} = 1$, $\tilde c^{(1)} = \tilde q-1 =
q-s-1$, $\tilde c^{(2)} = \tilde q^{\, 2}-3\tilde q + 1$, etc.  For M\"obius
strips, the switching of certain $\tilde c^{(d)}$'s, as specified for $s=1$ in
general in \cite{cf,zth}, generalizes to arbitrary $s$ in the interval $0 \le s
\le q$.  Further, from Eq. (\ref{nzhlyly}), it follows that there is only one
term with $d=L_y$, and we find (dropping the $j$ subscript)
\beq
\lambda_{Z,L_y,L_y}(q,s,v,w) = v \ . 
\label{lambdalyly}
\eeq
For $Ph(G_s, L_y \times m,cyc.,q,s,w) = Z(G, L_y \times m,cyc.,q,s,-1,w)$, we
have
\beq
Ph(G, L_y \times m,cyc.,q,s,w) = \sum_{d=0}^{L_y} \tilde c^{(d)} 
\sum_{j=1}^{n_{Ph}(L_y,d,s)} [\lambda_{Ph,G_s,L_y,d,j}(q,s,w)]^m 
\label{phsumcyc}
\eeq
As with $Z$, there is only one term with $d=L_y$,
and Eq. (\ref{lambdalyly}) shows that this is 
\beq
\lambda_{Ph,G_s,L_y,L_y}(q,s,w) = -1 \ . 
\label{phlambdalyly}
\eeq

The $n_{Zh}(L_y,d,s)$ satisfy the identity 
\beq
\sum_{d=0}^{L_y} \tilde c^{(d)} n_{Zh}(L_y,d,s) = q^{L_y}
\label{nzheq}
\eeq
while the $n_{Ph}(L_y,d,s)$ satisfy the identity
\beq
\sum_{d=0}^{L_y} \tilde c^{(d)} n_{Ph}(L_y,d) = P(L_y,q)=q(q-1)^{L_y-1} \ . 
\label{npheq}
\eeq
The reason why these identities hold for general $s$ with the same right-hand
side as for $s=0$ and $s=1$ is that the basic coloring constraints remain the
same; the only thing that is different for nonzero $s$ is the weighting
factors.  One method of calculating the $n_{Zh}(L_y,d,s)$ and $n_{Ph}(L_y,d,s)$
is to differentiate these respective equations $L_y$ times.  One thereby
obtains two respective sets of $L_y+1$ linear equations in the $L_y+1$ unknowns
$n_{Zh}(L_y,d,s)$ and $n_{Ph}(L_y,d,s)$ for $d=0,1,...,L_y$.  Solving these
equations determines these numbers $n_{Zh}(L_y,d,s)$ and $n_{Ph}(L_y,d,s)$.

We have used the method above to calculate the $n_{Zh}(L_y,d,s)$ and
$n_{Ph}(L_y,d,s)$.  For the $n_{Zh}(L_y,d,s)$ we find
\beq
n_{Zh}(L_y,L_y,s)=1
\label{nzhlyly}
\eeq
\beq
n_{Zh}(L_y,L_y-1,s)=(s+1)L_y+(L_y-1)
\label{nzhlylyminus1}
\eeq
\beq
n_{Zh}(L_y+1,0,s)=(s+1)n_{Zh}(L_y,0,s)+n_{Zh}(L_y,1,s)
\label{nzhlyd0}
\eeq
and, for $1 \le d \le L_y+1$, 
\beq
n_{Zh}(L_y+1,d,s)=n_{Zh}(L_y+1,d-1,s)+(s+2)n_{Zh}(L_y,d,s)+n_{Zh}(L_y,d+1,s)
\ . 
\label{nzhlyd}
\eeq
Some additional results, besides the general formulas for $n_{Zh}(L_y,d,s)$
with $d=L_y$ and $d=L_y-1$ given in Eqs. (\ref{nzhlyly}) and 
(\ref{nzhlylyminus1}), are 
\beq
L_y=2: \quad n_{Zh}(2,0,s)=s^2+2s+2
\label{nhly2numbers}
\eeq
\beqs
L_y=3: \quad & & n_{Zh}(3,0,s)=s^3+3s^2+6s+5, \cr\cr
             & & n_{Zh}(3,1,s)=3(s^2+3s+3)
\label{nhly1numbers}
\eeqs
\beqs
L_y=4: \quad & & n_{Zh}(4,0,s)=s^4+4s^3+12s^2+20s+14 \ , \cr\cr
             & & n_{Zh}(4,1,s)=4s^3+18s^2+36s+28 \ , \cr\cr
             & & n_{Zh}(4,2,s)=6s^2+20s+20 \ . 
\label{nzhly4numbers}
\eeqs

The numbers $n_{Zh}(L_y,d,s)$ of $\lambda_{Z,L_y,d,j}(q,s,v,w)$'s corresponding
to each $\tilde c^{(d)}$ in the general Potts model partition function are
reduced for the special case $v=-1$ that yields the weighted-set chromatic
polynomial.  By coloring combinatoric arguments similar to those used in
\cite{cf} and \cite{zth}, we determine the $n_{Ph}(L_y,d,s)$ as follows.  The
numbers $n_{Ph}(L_y,d,s)$ are identically zero for $d > L_y$, and
\beq
n_{Ph}(L_y,L_y,s)=1
\label{nphlydly}
\eeq
\beq
n_{Ph}(L_y,L_y-1,s)=(s+1)L_y
\label{nphlydlym1}
\eeq
\beq
n_{Ph}(L_y+1,0,s)=s \, n_{Ph}(L_y,0,s)+n_{Ph}(L_y,1,s)
\label{nphlyd0}
\eeq
and, for $1 \le d \le L_y+1$, 
\beq
n_{Ph}(L_y+1,d,s)=n_{Ph}(L_y+1,d-1,s)+(s+1)n_{Ph}(L_y,d,s)+n_{Ph}(L_y,d+1,s)
 \ . 
\label{nphlyd}
\eeq
Some additional results, besides the general formulas for $n_{Ph}(L_y,d,s)$ for
$d=L_y$ and $d=L_y-1$ given in Eqs. (\ref{nphlydly}) and 
(\ref{nphlydlym1}), are 
\beq
L_y=2: \quad n_{Ph}(2,0,s)=s^2+s+1
\label{nphly2numbers}
\eeq
\beqs
L_y=3: \quad & & n_{Ph}(3,0,s)=s^3+s^2+3s+2 \ , \cr\cr
             & & n_{Ph}(3,1,s)=3s^2+5s+4
\label{nphly3numbers}
\eeqs
\beqs
L_y=4: \quad & & n_{Ph}(4,0,s)=s^4+s^3+6s^2+7s+4 \ , \cr\cr
             & & n_{Ph}(4,1,s)=4s^3+9s^2+15s+9 \ , \cr\cr
             & & n_{Ph}(4,2,s)=6s^2+11s+8 \ . 
\label{nphly4numbers}
\eeqs
As we have noted above, for $s=0$, $Z(G,q,s,v,w)$ reduces to $Z(G,q,v)$, so we
focus on the nonzero (integer) $s$ values in the set $I_s$.  For these values
we find 
\beq
n_{Ph}(L_y,d,s) = n_{Zh}(L_y,d,s-1) + n_{Zh}(L_y-1,d,s-1) \ . 
\label{nphzlyd_relation}
\eeq

From our determination of the numbers $n_{Zh}(L_y,d,s)$ and 
$n_{Ph}(L_y,d,s)$, we next calculate the total numbers
\beq
N_{Zh,L_y,s} = \sum_{d=0}^{L_y} n_{Zh}(L_y,d,s) 
\label{nzhlytotsum}
\eeq
and
\beq
N_{Ph,L_y,s} = \sum_{d=0}^{L_y} n_{Ph}(L_y,d,s) \ . 
\label{nphlytotsum}
\eeq
From Eq. (\ref{nphzlyd_relation}) it follows that for nonzero (integer) $s$
values in the set $I_s$, 
\beq
N_{Ph,L_y,s} = N_{Zh,L_y,s-1} + N_{Zh,L_y-1,s-1} \ . 
\label{nphlytot_relation}
\eeq
Using our results in Eqs. (\ref{nzhlyly})-(\ref{nzhlyd}), we find that, for 
nonzero $s \in I_s$, 
\beq
N_{Zh,L_y,s}=\sum_{j=0}^{L_y} {L_y \choose j} \, {2j \choose j} \, s^{L_y-j}
\label{nzlytots}
\eeq
and hence, using Eq. (\ref{nphlytot_relation}), we have 
\beq
N_{Ph,L_y,s}=\sum_{j=0}^{L_y} {L_y \choose j} \, {2j \choose j} \, 
(s-1)^{L_y-j} +
\sum_{j=0}^{L_y-1} {L_y-1 \choose j} \, {2j \choose j} \, (s-1)^{L_y-1-j} \ . 
\label{nphlytots}
\eeq
A few explicit results for low values of $s$ are 
\beq
N_{Zh,1,s}=s+2
\label{nzly1tots}
\eeq
\beq
N_{Zh,2,s}=s^2+4s+6
\label{nzly2tots}
\eeq
\beq
N_{Zh,3,s}=s^3+6s^2+18s+20=(s+2)(s^2+4s+10)
\label{nzly3tots}
\eeq
\beq
N_{Zh,4,s}=s^4+8s^3+36s^2+80s+70
\label{nzly4tots}
\eeq
\beqs
N_{Zh,5,s} & = & s^5+10s^4+60s^3+200s^2+350s+252 \cr\cr
           & = & (s+2)(s^4+8s^3+44s^2+112s+126)
\label{nzly5tots}
\eeqs
\beq
N_{Zh,6,s}=s^6+12s^5+90s^4+400s^3+1050s^2+1512s+924
\label{nzly6tot}
\eeq
and
\beq
N_{Ph,1,s}=s+2
\label{nphly1tots}
\eeq
\beq
N_{Ph,2,s}=s^2+3s+4
\label{nphly2tots}
\eeq
\beq
N_{Ph,3,s}=s^3+4s^2+11s+10
\label{nphly3tots}
\eeq
\beq
N_{Ph,4,s}=s^4+5s^3+21s^2+37s+26
\label{nphly4tots}
\eeq
\beq
N_{Ph,5,s}=s^5+6s^4+34s^3+88s^2+123s+70
\label{nphly5tots}
\eeq
and
\beq
N_{Ph,6,s}=s^6+7s^5+50s^4+170s^3+366s^2+401s+192 \ . 
\label{nphly6tots}
\eeq

For $L_y >> 1$, these total numbers have the following 
dominant asymptotic exponential growth rates (suppressing power-law 
prefactors): 
\beq
N_{Zh,L_y,s} = (s+4)^{L_y} \quad {\rm for} \ L_y \to \infty
\label{nzhlytot_asymp}
\eeq
and
\beq
N_{Ph,L_y,s} = (s+3)^{L_y} \quad {\rm for} \ L_y \to \infty \ . 
\label{nphlytot_asymp}
\eeq

We note that the $s \to q-s$ symmetry (\ref{zsym}) is not manifestly evident in
the various results that we have given for $n_{Zh}(L_y,d,s)$,
$n_{Ph}(L_y,d,s)$, $N_{Zh,L_y,s}$, and $N_{Ph,L_y,s}$. This symmetry arises via
identities involving the $\tilde c^{(d)}$ and the $\lambda$'s.  The symmetry
(\ref{zsym}) implies that (for $s \in I_s$ and denoting $L_x \equiv m$)
\beqs
& & \sum_{d=0}^{L_y} c^{(d)}(q-s) \sum_{j=1}^{n_{Zh}(L_y,d,s)} 
[\lambda_{Z,L_y,d,j}(q,s,v,w)]^m = \cr\cr
& & 
 w^n\sum_{d=0}^{L_y} c^{(d)}(s) \sum_{j=1}^{n_{Zh}(L_y,d,q-s)} 
[\lambda_{Z,L_y,d,j}(q,q-s,v,w^{-1})]^m
\label{zsumsym}
\eeqs
and, for $v=-1$, 
\beqs
& & \sum_{d=0}^{L_y} c^{(d)}(q-s) \sum_{j=1}^{n_{Ph}(L_y,d,s)} 
[\lambda_{Ph,L_y,d,j}(q,s,w)]^m = \cr\cr
& & 
 w^n\sum_{d=0}^{L_y} c^{(d)}(s) \sum_{j=1}^{n_{Ph}(L_y,d,q-s)}
[\lambda_{Ph,L_y,d,j}(q,q-s,w^{-1})]^m \ . 
\label{phsumsym}
\eeqs
In particular, when $s=q$, $Z$ simplifies to a multiple times the zero-field
$Z$ as specified by the general relation (\ref{zsq}), with a consequent
reduction in the number of distinct $\lambda_{Z,L_y,d,j}(q,s,v,w)$'s.  Using
the fact that $c^{(d)}(0)=(-1)^d$ \cite{cf}, we can express this reduction as
\beqs
& & \sum_{d=0}^{L_y} (-1)^d \sum_{j=1}^{n_{Zh}(L_y,d,q)} 
[\lambda_{Z,L_y,d,j}(q,q,v,w)]^m = \cr\cr
& & w^n\sum_{d=0}^{L_y} c^{(d)}(q) \sum_{j=1}^{n_Z(L_y,d)} 
[\lambda_{Z,L_y,d,j}(q,v)]^m
\label{zsumsymsq}
\eeqs
where $n_Z(L_y,d) \equiv n_{Zh}(L_y,d,0)$.  For $s=q-1$, Eq. (\ref{zsumsym})
yields the identity 
\beqs
& & \sum_{d=0}^{L_y} c^{(d)}(1) \sum_{j=1}^{n_{Zh}(L_y,d,q-1)} 
[\lambda_{Z,L_y,d,j}(q,q-1,v,w)]^m = \cr\cr
& & w^n\sum_{d=0}^{L_y} c^{(d)}(q-1) \sum_{j=1}^{n_{Zh}(L_y,d,1)} 
[\lambda_{Z,L_y,d,j}(q,1,v,w^{-1})]^m
\label{zsumsymsqminus1}
\eeqs
where $c^{(d)}(1)$ takes on the values \cite{cf}
\beq
c^{(d)}(1)= \cases{ 1 & if $d=0$ \ mod \ 3 \cr
                    0 & if $d=1$ \ mod \ 3 \cr
                   -1 & if $d=2$ \ mod \ 3  } \ . 
\label{cdq1}
\eeq
Analogous identities follow from Eq. (\ref{zsumsym}) for $s=q-2$, $s=q-3$
and $s=q-4$ (assuming that $s \in I_s$), where the values of $c^{(d)}(2)$,
$c^{(d)}(3)$, and $c^{(d)}(4)$ were given in Eqs. (2.19)-(2.21) of
Ref. \cite{cf}.  Thus, in using the results above for $n_{Zh}(L_y,d,s)$,
$n_{Ph}(L_y,d,s)$, $N_{Zh,L_y,s}$, and $N_{Ph,L_y,s}$, it is understood that
they apply for generic values of $s \in I_s$ but involve simplifications for
special values of $s$. We shall give an example of this in the next section.

\section{Circuit Graphs $C_n$}

The circuit graph $C_n$, or equivalently, the 1D lattice with periodic boundary
conditions, has chromatic number
\beq
\chi(C_n) = \cases{ 2 & if $n \ge 2$ \ is \ even \cr
                    3 & if $n \ge 3$ \ is \ odd } 
\label{chicn}
\eeq
(The case $n=1$ is a single vertex with a loop, for which
there is no proper $q$-coloring, so $Ph(C_1,q,w)$ vanishes identically.) 
In general, $Z(C_n,q,s,v,w)$ has the structure 
\beq
Z(C_n,q,s,v,w) = \sum_{j=1}^{s+1} [\lambda_{Z,1,0,j}(q,s,v,w)]^m + 
\tilde c^{(1)} v^m 
\label{zcn}
\eeq
where we recall that $\tilde c^{(1)}=q-s-1$.  We find that 
(suppressing arguments) the $\lambda_{Z,1,0,j}$ for $1 \le j \le s+1$ are 
given by
\beqs
\lambda_{Z,1,0,j} & = & \frac{1}{2} \Bigg [ q-s+v + w(s+v) \pm 
 \Big [ \{q-s+v + w(s+v)\}^2-4vw(q+v)\Big ]^{1/2} \ \Bigg ] 
\label{lamcnplusminus}
\eeqs
for $j=1,2$ (which is the total set for $s=0$ or $s=1$), and, for $s \ge 2$, 
\beq
\lambda_{Z,1,0,j} = vw \ \  {\rm for} \ \ 3 \le j \le s+1
\label{lamvw}
\eeq
That is, for general $s \in I_s$, 
\beq
Z(C_n,q,s,v,w) = \sum_{j=1}^2 [\lambda_{Z,1,0,j}]^n + (s-1)(vw)^n + (q-s-1)v^n
\ . 
\label{zcnexplicit}
\eeq
It is readily checked that this expression for
$Z(C_n,q,s,v,w)$ (i) reduces to the zero-field Potts model partition function 
\beq
Z(C_n,q,v) = (q+v)^n + (q-1)v^n
\label{zcnh0}
\eeq
for $s=0$ or $w=1$, (ii) satisfies the general symmetry property
(\ref{zsym}), and (iii) reduces to $w^nZ(C_n,q,v)$ for $s=q$, in agreement with
Eq. (\ref{zsq}).  

As was noted briefly at the beginning of the paper, our result
(\ref{zcnexplicit}) shows a qualitative difference between the case $s=1$
considered previously \cite{ph} and the more general set of cases with $s \ge
2$ in the interval $I_s$, namely the fact that the third term, $(s-1)(vw)^n$,
is absent for $s=1$ but is present for other values of $s \in I_s$.  By the $s
\leftrightarrow q-s$ symmetry inherent in Eqs. (\ref{symgroup}) and
(\ref{zsym}), this also means that another term vanishes identically for
$s=q-1$ but is present for other values of $s \in I_s$; this is the last term
in (\ref{zcnexplicit}), $(q-s-1)v^n$.  It is interesting to observe that the
symmetry (\ref{zsym}) applies not just to the total $Z(C_n,q,s,v,w)$, but also
to parts of this function.  Specifically, under the replacement $s \to q-s$,
one sees that (i) the sum of the last two terms in (\ref{zcnexplicit}),
$(s-1)(vw)^n + (q-s-1)v^n$, transforms into $w^n[ (q-s-1)v^n+(s-1)(vw^{-1})^n]$
and (ii) the first two terms, $\sum_{j=1}^2 [\lambda_{Z,1,0,j}(q,s,v,w)]^n$
transform into $w^n\sum_{j=1}^2 [\lambda_{Z,1,0,j}(q,q-s,v,w^{-1})]^n$, so that
each of these parts, (i) and (ii), individually satisfies the symmetry
(\ref{zsym}).

We exhibit $Z(C_n,q,s,v,w)$ for $2 \le n \le 4$ below.  To keep the equations
as compact as possible, we write the coefficients of the terms of maximal
degree in $w$ and of degree 0 in $w$ in terms of zero-field partition functions
using the general results (\ref{zbetajn}) and (\ref{zbetaj0}). We find 
\beqs
Z(C_2,q,s,v,w) & = &Z(C_2,s,v)w^2 + 2s(q-s)w + Z(C_2,q-s,v) \cr\cr
               & = & q^2+[2t+v(v+2)]q+t \Big [t+v(v+2)(w+1) \Big ]
\label{zc2}
\eeqs
\beqs
Z(C_3,q,s,v,w) & = & Z(C_3,s,v)w^3 + 3s(q-s)(s+v)w^2 \cr\cr
               & + & 3s(q-s)(q-s+v)w+Z(C_3,q-s,v)
\label{zc3}
\eeqs
\beqs
Z(C_4,q,s,v,w) & = & Z(C_4,s,v) w^4 +4s(q-s)(s+v)^2 w^3 \cr\cr
               & + & 2s(q-s)\Big [3(q^2-s^2-(q-s)^2)+4qv+4v^2\Big ] w^2 \cr\cr
               & + & 4q(q-s)(q-s+v)^2 w + Z(C_4,q-s,v)  \ . 
\label{zc4}
\eeqs
As usual, one obtains the $Ph(C_n,q,s,w)$ for each $n$ by setting $v=-1$ in
$Z(C_n,q,s,v,w)$. For $s=1$, the parts of $Z(C_n,q,1,v,w)$ were given in
Ref. \cite{zth} and $Ph(C_n,q,1,w)$ was given in Ref. \cite{ph}.  

In the context of weighted-set coloring, so that $s$ is an integer in the
interval $I_s$, it was noted \cite{ph} that for $s=1$, $Ph(C_n,q,1,w)$ contains
the factor $(q-1)$.  It was also noted that if $n$ is odd, say $n=2m+1$ with
$m=1,2,..$ then $Ph(C_{2m+1},q,1,w)$ also contains the factor $(q-2)$, so that
for odd $n$, $Ph(C_{2m+1},2,1,w)=0$ \cite{ph}.  The fact that if $n=2m+1$ is
odd and $q=2$, then one cannot perform a proper $q$-coloring of $C_n$, so that
$Ph(C_{2m+1},2,s,w)=0$, is independent of both $w$ and $s$.  However, in
contrast to the $s=1$ case, where the vanishing of $Ph(C_{2m+1},q,1,w)$ for
$q=2$ occurred via a factor of $(q-2)$, this is not the case for general $s$.
Instead, $Ph(C_{2m+1},q,s,w)$ is such that if one evaluates it at $q=2$, there
is a factor which is a polynomial in $s$ that implicitly but necessarily
vanishes.  This vanishing occurs because of the implicit restriction on the
values of $s$, namely that $s$ is an integer in the interval $0 \le s \le q$.
For example, consider $Ph(C_3,q,s,w)$, given above as the $v=-1$ special case
of Eq. (\ref{zc3}). Evaluating this at $q=2$, we obtain
\beq
Ph(C_3,2,s,w) = s(s-1)(s-2)(w-1)^3 \ . 
\label{phc3q2}
\eeq
This implicitly vanishes for any of the allowed (integer) values of $s$ in the
set $I_s$ because for $q=2$, $s$ can only take on the values 0, 1, or 2 in this
set. The same type of mechanism is responsible for the vanishing of
$Ph(C_{2m+1},q,s,w)$ at $q=2$ for higher values of $m$.  For $s=2$, if $n \ge
3$ is odd, then our results show that $Ph(C_n,q,s,w)$ contains the factor
$(q-2)$.  If $s \ge 3$, then, in general, $Ph(C_n,q,s,w)$ does not have such
overall factors.  The reason for this is that $s$ is bounded above by $q$, so
that if $s \ge 3$, then $q \ge 3$.  This is equal to $\chi(C_n)$ for $n$ odd
and greater than $\chi(C_n)$ for $n$ even (cf. Eq. (\ref{chicn})), so that 
$Ph(C_n,q,s,w)$ is not forced to vanish the way it is for the cases of $s=0, \
1, \ 2$ when $q$ can be less then $\chi(C_n)$.  

We have mentioned above that our general structural formulas for $Z(G,q,s,v,w)$
and $Ph(G,q,s,w)$ with $G$ a cyclic strip graph simplify considerably when
$s=q$, as required by the general relations (\ref{zsq}) and (\ref{phsq}).  It
is interesting to see how this occurs for this family of circuit graphs,
$G=C_n$.  There are $N_{Zh,L_y,s}=s+2$ terms $\lambda_{Z,L_y,d,j}(q,s,v,w)$
with $L_y=1$ whose $n$'th powers occur in Eq. (\ref{zcn}).  Of these,
$n_{Zh,1,0,s}=s+1$ multiply the coefficient $\tilde c^{(0)}=1$ and the
remaining term, $[\lambda_{Z,1,1}(q,s,v,w)]^n = v^n$, multiplies the
coefficient $\tilde c^{(1)}=q-s-1$.  From the general relation (\ref{zsq}), we
know that for $s=q$, $Z(C_n,q,q,v,w)=w^n \, Z(C_n,q,v)$, where $Z(C_n,q,v)$ is
the zero-field Potts model partition function for the circuit graph $C_n$,
given by Eq. (\ref{zcnh0}). Hence, we can deduce that of the $s+1$ terms
$\lambda_{Z,1,0,j}(q,s,v,w)$, (i) one becomes equal to $w(q+v)$; (ii) a second
becomes equal to $v$, and (iii) the remaining $s-1=q-1$ terms become equal to
$wv$. Using the fact that $s=q$, so that $\tilde c^{(1)}=-1$, we then have the
reduction
\beqs
Z(C_n,q,q,v,w) & = & \sum_{j=1}^{s+1} [\lambda_{Z,1,0,j}(q,q,v,w)]^n 
+ \tilde c^{(1)} v^n \cr\cr
               & = & [w(q+v)]^n + v^n + (q-1)(wv)^n - v^n \cr\cr
               & = & w^n \Big [ (q+v)^n + (q-1)v^n \Big ] \cr\cr
               & = & w^n \, Z(C_n,q,v) \ . 
\label{zcnsqreduction}
\eeqs
Thus, the identity (\ref{zsq}) (a special case of the symmetry (\ref{zsym})) is
realized via a ``transmigration'' process in which one or more
$\lambda_{Z,L_y,d,j}(q,s,v,w)$'s for a given $d$ become equal or proportional
to (respectively, one or more) $\lambda_{Z,L_y,d',j}(q,s,v,w)$'s for a
different $d'$ and hence their $m$'th powers can be regrouped with the latter,
thereby changing the effective coefficients that multiply the
$[\lambda_{Z,L_y,d',j}(q,s,v,w)]^m$'s , where here $m=n$.  This transmigration
process is a general one and occurs also for higher values of strip width
$L_y$.  This process is also the mechanism whereby the identities (\ref{zsym})
and (\ref{phsym}) are satisfied.  Thus, the results above for
$n_{Zh}(L_y,d,s)$, $n_{Ph}(L_y,d,s)$, $N_{Zh,L_y,s}$, and $N_{Ph,L_y,s}$ apply
for generic values of $s$, but there are simplifications, involving this type
of transmigration, for special values of $s$.

\section{Some Properties of the Zeros of $Ph(G,q,s,w)$} 

One can solve for the zeros of $Z(G,q,s,v,w)$ as functions 
of any of the four variables $q$, $s$, $v$ and $w$ with the other three
held fixed, and similarly, one can solve for the zeros of $Ph(G,q,s,w)$ as a
function of any of the three variables $q$, $s$, and $w$ with the other two
held fixed.  Here it is understood that one uses Eq. (\ref{clusterws}), with
these four variables each generalized to lie in ${\mathbb C}$. 

In general, some zeros in $w$ and $s$ can have unbounded magnitudes as
a function of the other variables.  The underlying reason for this can be seen
at an algebraic level as the fact that the coefficient of the highest-degree
term in this variable in $Z(G,q,s,v,w)$ can vanish at some values of $w$ 
and/or $s$.  Related to this are the facts that (i) at the special values 
$w=1$ and $w=0$, $Z(G,q,s,v,w)$ loses its dependence on $s$, and (ii) at the
special value $s=0$, $Z(G,q,s,v,w)$ loses its dependence on $w$.  As a
consequence, the zeros in these variables move off to infinity. Similar
statements apply for $Ph(G,q,s,w)$.  We proceed to discuss some properties of
these zeros.

\subsection{Zeros of $Z(G,q,s,v,w)$ and $Ph(G,q,s,w)$ in $q$}

Here we consider the zeros of $Z(G,q,s,v,w)$ and $Ph(G,q,w)$ in $q$, as a
function of $s$ and $w$, for some graphs $G$.  Since the maximal degree of
$Z(G,q,s,v,w)$ and $Ph(G,q,s,w)$ in the variable $q$ is $n(G)$, each of these
polynomials has this number of zeros in the variable $q$.  In general, since
$Z(G,q,s,v,1)=Z(G,q,v)$ while $Z(G,q,s,v,0)=Z(G,q-s,v)$, it follows that as $w$
decreases from 1 to 0, there is an overall shift to the right in the zeros of
$Z(G,q,s,v,w)$ in the $q$ plane by $s$ units, and this holds, in particular,
for the case $v=-1$ that yields $Ph(G,q,s,w)$.  This shift is illustrated by
some simple examples.  For $G=L_1=K_1=N_1$, $Z(L_1,q,s,v,w)=0$ at
\beq
q_{L1z}=-t=s(1-w) \ .  
\label{qzline1}
\eeq
This increases from 0 to $s$ as $w$ decreases from 1 to 0 in the DFSCP interval
and decreases from 0 through negative values as $w$ increases above 1 in the
FSCP interval. For $L_2=K_2$, $Z(L_2,q,s,v,w)$ vanishes at
\beq
q_{L2z,j} = \frac{1}{2}\left [ -v+2s(1-w) \pm \sqrt{v[v-4sw(w-1)]} \right ] 
\label{qzline2}
\eeq
where $j=1,2$ for the $\pm$ sign.  As $w$ decreases from 1 to 0, the root
$q_{L2z,1}$ increases from 0 to $s$, while the root $q_{L2z,2}$ increases from
$-v$ to $s-v$.  The zeros of graph-coloring polynomials in $q$ satisfy certain
boundedness properties for fixed nonzero magnetic field \cite{sokalbound}.
However, the zeros of $Z(G,q,s,v,w)$ and $Ph(G,q,s,w)$ in $q$ are, in general,
unbounded as $|w| \to \infty$.  This is already evident in the simplest case of
a single vertex, for which the zero of $Z(L_1,q,s,v,w)$ at $q=s(1-w)$ has a
magnitude that, for $s \ne 0$, goes to infinity as $|w| \to \infty$.

\subsection{Zeros of $Z(G,q,s,v,w)$ and $Ph(G,q,s,w)$ in $s$} 

One may also study the zeros of $Z(G,q,s,v,w)$ in each of the three variables
$s$, $v$, and $w$ with the other two (and $q$) held fixed, and the zeros of
$Ph(G,q,s,w)$ in each of the two variables $s$ and $w$ with the other (and $q$)
held fixed.  In general, these zeros are unbounded in magnitude even when the
other variables vary over a finite range.  The reason for the divergences in
these zeros of $Z(G,q,s,v,w)$ is the fact that the coefficient of the term in
$Z(G,q,s,v,w)$ of highest degree in the given variable can vanish as one
changes other variables, and similarly with $Ph(G,q,s,w)$. Again, this can be
illustrated with simple examples.  

We begin our discussion with the zeros in $s$, where it is understood that
this variable is formally extended from the integers in the interval $I_s$ to
the complex numbers.  The Potts model partition function for the single-vertex
graph $L_1$, $Z(L_1,q,s,v,w)$, vanishes at
\beq
s = \frac{q}{1-w} \ . 
\label{szline1}
\eeq
For $q \ne 0$, this diverges as $w \to 1$. For $L_2$, $Z(L_2,q,s,v,w)$ has
zeros in $s$ at
\beq
s_{L2z,j} = \frac{-[2q+v(w+1)] \pm \sqrt{v[v(w+1)^2+4qw]}}{2(w-1)}
\label{sline2zero}
\eeq
where $j=1,2$ correspond to the $\pm$ signs, respectively.  We find that 
\beq
s_{L2z,j} \sim \frac{-(q+v) \pm \sqrt{v(q+v)}}{w-1} \quad {\rm as} \ w \to 1
\label{sline2zerodiv}
\eeq
so that as $w \to 1$, the magnitudes $|s_{L2z,j}|$ are, in general, unbounded.
These results for the full Potts model partition function apply, {\it a
fortiori}, to the special case $v=-1$ that defines the weighted-set chromatic
polynomial, $Ph(L_2,q,s,w)$.  Above, we have explained the origin of this type
of divergence as being due to the fact that the coefficient of the
highest-power term of $Z(G,q,s,v,w)$ and $Ph(G,q,s,w)$ in the given variable,
here $s$, vanishes.  This occurs at $w=1$.  Closely related to this, at $w=1$,
all dependence on $s$ in $Z(G,q,s,v,w)$ disappears, so it is understandable
that the zeros in $s$ would disappear by moving off to infinity in this limit.

\subsection{Zeros of $Z(G,q,s,v,w)$ and $Ph(G,q,s,w)$ in $w$} 

Here we comment on zeros of $Z(G,q,s,v,w)$ and $Ph(G,q,s,w)$ in $w$.  First, we
note that the symmetry (\ref{zsym}) implies that if one replaces $s$ by $q-s$,
then the zeros of $Z(G,q,s,v,w)$ and $Ph(G,q,s,w)$ in $w$ away from the origin
map into their inverses. In particular, if $q=2s$, then Eq. (\ref{zsym})
reads $Z(G,2s,s,v,w)=w^nZ(G,2s,s,v,w^{-1})$, so that the zeros of $Z$ away from
the origin in the $w$ plane form a set that is invariant under inversion. 
For $G=L_1$, $Z(L_1,q,s,v,w)$ vanishes at 
\beq
w_{L1z} = 1 - \frac{q}{s} \ . 
\label{line1wzero}
\eeq
From Eq. (\ref{zline2}), we find that $Z(L_2,q,s,v,w)=0$ for
\beq
w_{L2z,j} = \frac{s(s-q) \pm \sqrt{s(s-q)v(q+v)}}{s(s+v)}
\label{line2wzero}
\eeq
where $j=1,2$ correspond to the $\pm$ signs, respectively.  The right-hand
sides of Eqs. (\ref{line1wzero}) and (\ref{line2wzero}) both diverge as $s \to
0$.  In particular, 
\beq
w_{L2z,j} \sim \pm \sqrt{\frac{(-q)(q+v)}{sv}} \quad {\rm as} \  s \to 0 
\label{wline2zeros_s0}
\eeq
for $j=1,2$.  The fact that, in general, the magnitudes of the zero $w_{L1z}$
of $Z(L_1,q,s,v,w)$ and the zeros $w_{L2z,j}$, $j=1,2$, of $Z(L_2,q,s,v,w)$
diverge as $s \to 0$ is again understandable, since for a graph $G$, if $s=0$,
then $Z(G,q,s,v,w)$ reduces to $Z(G,q,v)$, with no dependence on $w$.  Hence,
in the limit $s \to 0$, it is natural that the zeros in $w$ disappear by moving
off to infinity.  Finally, as $s \to -v$, we find that
\beq
w_{L2z,2} \to \frac{q+2v}{2v} \quad {\rm as} \  s \to -v 
\label{wline2zero_bnd}
\eeq
while $w_{L2z,1}$ is unbounded:
\beq
w_{L2z,1} \sim -\frac{2(q+v)}{s+v} \quad {\rm as} \  s \to -v \ . 
\label{wline2zero_smv}
\eeq

\subsection{Zeros of $Z(G,q,s,v,w)$ in $v$} 

To illustrate the calculation of the zeros of $Z(G,q,s,v,w)$ in $v$, we 
again use our simple example graph, $L_2$.  We find the zero of
$Z(L_2,q,s,v,w)$ in $v$ occurs at
\beq
v = -\frac{[q+s(w-1)]^2}{q+s(w-1)(w+1)} \ . 
\label{vline2zero}
\eeq
This has an unbounded magnitude if $q+s(w^2-1) = 0$, i.e.,
\beq
s = \frac{q}{1-w^2} \ . 
\label{sz}
\eeq
This divergence in the magnitude of the right-hand side of (\ref{vline2zero})
does not directly affect the weighted-set proper vertex coloring of this $L_2$
graph because in the DFSCP region $0 \le w < 1$, the condition (\ref{sz})
implies that $s > q$, outside the actual coloring interval $I_s$, and in the
FSCP region $w > 1$, it implies that $s$ is negative, again outside of this
interval $I_s$.

\section{Quantities Defined in the Limit $n(G) \to \infty$} 

\subsection{$f$ and $\Phi$ Functions} 

Let us consider families of graphs $G_m$ that can be built up recursively, such
as lattice strips.  For such graphs, the $m+1$ member of the family is obtained
from the $m$ th member by (possibly cutting and) gluing in a given subgraph.
For example, for the graph $C_m$, one cuts the circuit at some vertex and
inserts another edge and vertex to get $C_{m+1}$, and so forth. Generalizing
this, another example is a strip of the square lattice with transverse width
$L_y$, length $L_x=m$, and periodic longitudinal boundary conditions, which we
denote as $sq(L_y \times L_x,cyc.)$.  In this case, the number of vertices is
$n=L_yL_x$. Generically, the number of vertices of a recursive graph $G_m$ is
of the form $n(G_m)=am+b$, where $a$ and $b$ are (integer) constants, so that
the limit $n \to \infty$ is equivalent to the limit $m \to \infty$.  We denote
the formal $n \to \infty$ limit of such graphs as $\{G\} = \lim_{m \to \infty}
G_m$.  In the present context, this $n \to \infty$ limit corresponds to the
limit of infinite length for a strip graph of fixed width and some prescribed
boundary conditions.  Correspondingly, one can define the free energy per
vertex as follows (with the subscript $m$ suppressed in the notation) 
\beq
f(\{G\},q,s,v,w) = \lim_{n \to \infty} n^{-1} \ln [Z(G,q,s,v,w)]
\label{fdef}
\eeq
and a function $\Phi(\{G\},q,s,w)$, 
\beq
\Phi(\{G\},q,s,w) = \lim_{n \to \infty} [Ph(G,q,s,w)]^{1/n} \ . 
\label{phidef}
\eeq
As before (cf. Eq. (1.9) of \cite{w} and Eq. (2.8) of \cite{a}), one must take
account of a noncommutativity of limits that can occur, namely the fact that
for certain special values of $q$, denoted $\{ q_{sp} \}$, the limits $n \to
\infty$ and $q \to q_{sp}$ do not commute:
\beq
\lim_{n \to \infty} \lim_{q \to q_{sp}} Z(G,q,s,v,w)^{1/n} \ne
\lim_{q \to q_{sp}} \lim_{n \to \infty} Z(G,q,s,v,w)^{1/n} 
\label{znoncom}
\eeq
and the analogous formulas for the $v=-1$ case which defines
$\Phi(\{G\},q,s,w)$.  For further details, we refer the reader to our previous
discussions of this \cite{w,a,ph}.  An explicit example is provided by our
result for $Z(C_n,q,s,v,w)$ in Eq. (\ref{zcnexplicit}); if one sets $q=s+1$
first before taking $n \to \infty$, then the last term drops out, while if one
takes $n \to \infty$ first with $q \ne s+1$, then, since $\lim_{n \to \infty}
[(q-s-1)v^n]^{1/n} = v$, the last term may remain in $f$.  We see also an
additional type of noncommutativity that is present, namely that if we extend
$s$ from an integer in $I_s$ to a real (or complex) variable, then for a set of
special values of $s$, denoted $\{s_{sp}\}$, 
\beq
\lim_{n \to \infty} \lim_{s \to s_{sp}} Z(G,q,s,v,w)^{1/n} \ne
\lim_{s \to s_{sp}} \lim_{n \to \infty} Z(G,q,s,v,w)^{1/n} \ . 
\label{znoncom2}
\eeq
Again, our result for $Z(C_n,q,s,v,w)$ in Eq. (\ref{zcnexplicit}) provides an
illustration of this; if one sets $s=1$ first before taking $n \to \infty$,
then the third term, $(s-1)(vw)^n$, drops out, while if one takes $n \to
\infty$ first with $s \ne 1$, and then sets $s=1$, it follows that, since
$\lim_{n \to \infty} [(s-1)(vw)^n]^{1/n} = vw$, the resultant limiting 
term $vw$ may remain in $f$.
Similarly, if one sets $s=q-1$ before taking $n \to \infty$, then the last
term, $(q-s-1)v^n$ drops out, while if one takes $n \to \infty$ first with $s
\ne q-1$, and then sets $s=q-1$, it follows that since $\lim_{n \to \infty}
[(q-s-1)v^n]^{1/n} = v$, the resultant limiting term $v$ may remain in $f$.
In our analysis of the $n \to \infty$ limit for recursive families of graphs,
unless otherwise indicated, we shall choose the order of limits in which we fix
$s$ first and then take $n \to \infty$.

The function $Ph(\{G\},q,s,w)$ generalizes the ground state degeneracy of the
zero-field, zero-temperature Potts antiferromagnet, $W(\{G\},q) = \lim_{n \to
\infty} P(G,q)^{1/n}$. Thus (with care taken concerning the above-mentioned
noncommutativity of limits), $\Phi(\{G\},q,s,1) = \Phi(\{G\},q,0,w) =
W(\{G\},q)$.  In the case of the zero-field, zero-temperature Potts
antiferromagnet, the associated configurational entropy per vertex (which is
thus the ground-state entropy per site) $\{G\}$ is $S = k_B \ln W$.  The third
law of thermodynamics states that the entropy per site $S$ should vanish as the
temperature goes to zero.  However, there are a number of exceptions to this
law.  For instance, for the zero-field $q$-state Potts antiferromagnet on a
square lattice, an elementary argument yields the lower bound $S/k_B \ge
(1/2)\ln(q-1)$, which is nonzero for $q \ge 3$.

In the present case of the weighted-set chromatic polynomial, let us consider
first the FSCP interval $w > 1$ and assume that $q \ge 2$, so that a proper
$q$-coloring can be performed for a bipartite graph.  We also assume that $s$
is an integer for this discussion and lies in the interval $2 \le s \le
q$. Applying our lower bound (\ref{phboundlargew}) to a particular bipartite
graph, namely the strip graph of the square ($sq$) lattice with width $L_y$
vertices and even length $L_x$ vertices, denoted $sq(L_y \times L_x)$, we have
(with $n_1=n_2=L_xL_y/2$) the lower bound
\beq
Ph(sq(L_y \times L_x),q,s,w) \ge s(s-1)^{n/2} w^n  \ . 
\label{phstripbound}
\eeq
Taking $L_x \to \infty$ with $L_y$ fixed, we thus obtain the lower bound
$\Phi(sq(L_y \times \infty),q,s,w) \ge w\sqrt{s-1}$. 
Hence, in this limit, the entropy is bounded below by
\beq
S(sq(L_y \times \infty),q,s,w) \ge \ln w + \frac{1}{2}\ln(s-1)  \quad {(\rm 
FSCP \ case)} \ . 
\label{phboundlargewstrip}
\eeq
For $s \ge 3$ (which implies $q \ge 3$ also), this ground state entropy is
nonzero.  In the DFSCP interval $0 \le w < 1$, with $1 \le s \le q-3$, applying
our lower bound (\ref{phboundsmallw}) to the same limit of this strip graph, we
have $\Phi(sq(L_y \times \infty),q,s,w) \ge \sqrt{q-s-1}$, so that
\beq
S(sq(L_y \times \infty),q,s,w) \ge \frac{1}{2}\ln(q-s-1)  \quad {(\rm 
DFSCP \ case)} \ . 
\label{phboundsmallwstrip}
\eeq
For the given range, $q \ge s+3$, this entropy is again nonzero.

\subsection{Example of Calculation of $f$ and $\Phi$ for a Family of Graphs}

We illustrate the calculation of the functions $f$ and $\Phi$ for the $n \to
\infty$ limit of the circuit graph $C_n$, or equivalently, the one-dimensional
lattice with periodic boundary conditions.  We shall take $s$ to be a fixed
integer in the interval $0 \le s \le q$ for this analysis. For a given set of
values of $q$, $s$, $v$, and $w$, the functional form of $f$ is determined by
the term $\lambda_{Z,1,0,j}(q,s,v,w)$ in Eq. (\ref{zcnexplicit}) with the
largest magnitude.  For fixed $s$, $v$, and $w$ and sufficiently large real
$q$, this is $\lambda_{Z,1,0,1}$.  Following our nomenclature in earlier work
for $w=1$, we denote this region as region $R_1$.  As for the zero-field case,
$f$ and $\Phi$ in this region are the same for the $n \to \infty$ limit of the
line graph $L_n$ and the circuit graph $C_n$.  We denote these limits as
$\{L\}$ and $\{C\}$.  We thus have
\beq
f(\{L\},q,s,v,w) = f(\{C\},q,s,v,w) = \ln[\lambda_{Z,1,0,1}(q,s,v,w)]
\label{fcn}
\eeq
and 
\beq
\Phi(\{L\},q,s,w) = \Phi(\{C\},q,s,w) = \ln[\lambda_{Ph,1,0,1}(q,s,w)]
\label{phicn}
\eeq
where $\lambda_{Z,1,0,j}(q,s,v,w)$, $j=1,2$, were given for this family of
graphs in Eq. (\ref{lamcnplusminus}) and
$\lambda_{Ph,1,0,j}(q,s,w)=\lambda_{Z,1,0,j}(q,s,-1,w)$.  

It is of interest to consider how $\Phi(G,q,s,w)$ for $\{G\}=\{L\}$ (or
equivalently $\{G\}=\{C\}$ in region $R_1$) behaves for certain special cases
or limits of its variables.  For example, in the limit where the weighting is
removed, i.e., for $w \to 1$, $\Phi(\{L\},q,s,w)$ has the Taylor series
expansion
\beq
\Phi(\{L\},q,s,w) = q-1 + \frac{s(q-1)(w-1)}{q} -
\frac{s(q-1)(q-s)(w-1)^2}{q^3} + O\bigg ( (w-1)^3 \bigg ) \quad {\rm as} \ w
\to 1 \ . 
\label{phiw1}
\eeq
For $s=1$, it was shown in Ref. \cite{ph} that the
asymptotic behavior of $\Phi(\{L\},q,s,w)$ for large $|w|$ is
\beq
\Phi(\{L\},q,s,w) \sim \sqrt{(q-1)w} 
\left [ 1 + O \left ( \frac{1}{\sqrt{w}} \right ) \right ]  \quad {\rm for} \ 
 s=1 \ {\rm and} \ |w| \to \infty \ . 
\label{phi_s1_largew}
\eeq
In contrast, for $s \ne 1$, we find 
\beq
\Phi(\{L\},q,s,w) \sim (s-1)w + \frac{q-s-1}{2} + 
\left [ \frac{s(q-s)+q-1}{2(s-1)}\right ] + O \left ( \frac{1}{w} \right ) \ .
\label{phi_gen_largew}
\eeq
As is evident, the large-$|w|$ behavior is different, depending on
whether or not $s=1$.  For $|q| \to \infty$, we obtain the asymptotic expansion
\beq
\Phi(\{L\},q,s,w) \sim q+s(w-1)-1-\frac{sw(w-1)}{q} + 
O\left ( \frac{1}{q^2} \right ) \ . 
\label{phi_largeq}
\eeq
Extending $s$ from an integer in the interval $I_s$ to a real (or complex)
number, it is of interest to determine the limiting behavior of
$\Phi(\{L\},q,s,w)$ as $s \to 0$ and $s \to q$.  We calculate
\beq
\Phi(\{L\},q,s,w) = q-1 + \frac{(w-1)(q-1)s}{w+q-1} + O(s^2) 
\quad {\rm as} \ s \to 0
\label{phi_small_s}
\eeq
and
\beq
\Phi(\{L\},q,s,w) = w(q-1) - \frac{w(w-1)(q-1)(q-s)}{w(q-1)+1} + 
O \left ( (q-s)^2 \right ) \quad {\rm as} \ s \to q
\label{phi_qtos}
\eeq

From our study of $\Phi(\{G\},q,s,w)$ for the $n \to \infty$ limits of various
families of graphs, we have observed several generic properties.  Let us
consider families of strip graphs $G$ of regular lattices $\Lambda$.  Let the
chromatic number of the lattice $\Lambda$ be denoted as $\chi(\Lambda)$ and
assume that $q \ge \chi(\Lambda)$. A technical assumption is that the $n \to
\infty$ limit of the lattice strip graphs is taken in a manner such that for
each $G$, $\chi(G)=\chi(\Lambda)$.  (For example, for square-lattice strip
graphs with periodic longitudinal boundary, this means taking the length to be
even.)  Then for fixed $s \in I_s$ we have observed that (i) for fixed $w > 0$,
$\Phi(\{G\},q,s,w)$ is a monotonically increasing function of $q$ and (ii) for
fixed $q$, $\Phi(\{G\},q,s,w)$ is a monotonically increasing function of $w$
for $w > 0$.  One can also analyze the behavior of $\Phi(\{G\},q,s,w)$ as a
function of $s$ for fixed $q$ and $w$, but because of the noncommutativity
(\ref{znoncom2}), one must take care in specifying the order of limits used in
defining this function. The order that we take is first to set $s$ to a given
value in $I_s$ and then to take $n \to \infty$.  As is evident in the
definition of $I_s$, it is understood that $s \le q$.  We find that if $w > 1$,
then $\Phi(\{G\},q,s,w)$ is an increasing function of $s \in I_s$, while if $0
\le w < 1$, then $\Phi(\{G\},q,s,w)$ is a decreasing function of $s \in I_s$.

Focusing on the circuit graph, one sees that for values of the variables such
that another term $\lambda$ in Eq. (\ref{zcnexplicit}) becomes dominant, there
is a non-analytic change in $f$ and $\Phi$. As we have discussed earlier, this
is also associated with a locus, denoted generically $\cal B$, that comprises
the accumulation set of zeros of the respective function, $f$ and $\Phi$, For
definiteness, we analyze the locus $\cal B$ in the $q$ plane (denoted ${\cal
B}_q$) for $\Phi(\{C\},q,s,w)$, with $s$ and $w$ fixed. It may be recalled that
for the $n \to \infty$ limit of the unweighted chromatic polynomial, ${\cal
B}_q$ is the unit circle $|q-1|=1$, so $q_c=2$ in that case \cite{w,wc}.  For
the weighted-set chromatic polynomial, $Ph(C_n,q,s,w)$, the accumulation locus
${\cal B}$ depends on the value of $s$ and on whether $w$ is in the DFSCP
interval $0 \le w < 1$ or the FSCP interval $w > 1$.  We consider here the
DFSCP interval, since as $w$ decreases from 1 to 0, $Ph(C_n,q,s,w)$
interpolates between $P(G,q)$ and $P(G,q-s)$.  We also generally take $s \ne
0$, since for $s=0$ $Ph(C_n,q,s,w)$ reduces to the well-studied unweighted
chromatic polynomial $P(C_n,q)$.  (However, our results subsume this $s=0$
case.)

In this DFSCP interval, for $s=1$ or $s=2$, the lower boundary of the region
$R_1$ on the real axis, denoted $q_c$, is determined by the equality in
magnitude
\beq
|\lambda_{Ph,1,0,1}|=|\lambda_{Ph,1,1}|=1 \ ,
\label{lameqcn}
\eeq
which yields the result
\beq
q_c = 2 + \frac{s(1-w)}{1+w} \quad {\rm for} \ \{G\}=\{C\} \ {\rm and} \ 
 s=1 \ \ {\rm or} \ s=2 \ \ {\rm and} \ 0 \le w \le 1  \ . 
\label{qccn_wsmall}
\eeq
For $0 \le w < 1$, this value of $q_c$ is greater than the value $q_c=2$
for the unweighted chromatic polynomial.  Furthermore, as is evident from
Eq. (\ref{qccn_wsmall}), $q_c$ is a monotonically increasing function of $s$
for fixed $w$ in this DFSCP interval and a monotonically decreasing function of
$w$ for the fixed values of $s$ given above. As $w$ decreases from 1 to 0,
$q_c$ increases continuously from 2 to $2+s$.  In contrast, the left-hand part
of the boundary changes discontinuously; as $w$ decreases by an arbitrarily
small amount below 1, the point on the left where ${\cal B}_q$ crosses the real
$q$ axis jumps discontinuously from $q=0$ to $q=s$.  These results are in
accord with the fact that for $w=0$, ${\cal B}_q$ is the locus of solutions to
the equation $|q-(1+s)|=1$, i.e., the unit circle in the $q$ plane centered at
the point $q=1+s$, crossing the real axis on the left at $q=s$ and on the right
at $q=2+s$.  The locus ${\cal B}_q$ separates the $q$ plane into two regions.
We label the regions outside and inside the closed curve ${\cal B}$ 
as $R_1$ (noted before) and $R_2$, respectively. 

For $s > 2$, there is a change in the locus ${\cal B}_q$, because the
condition of degeneracy of leading $\lambda$'s is different; rather than 
Eq. (\ref{lameqcn}), it takes the form
\beq
|\lambda_{Ph,1,0,1}|=|\lambda_{Ph,1,0,2}| \ . 
\label{lameq2cn}
\eeq
This entails the condition that (i) 
\beq
q-s-1+w(s-1)=0 \ , 
\label{eqcnapart}
\eeq
so that $\lambda_{Ph,1,0,1}=-\lambda_{Ph,1,0,2}$, and the condition that (ii)
$|\lambda_{Ph,1,0,1}|=|\lambda_{Ph,1,0,2}| > 1$, so that these $\lambda$'s are
dominant. Substituting for $q$ from Eq. (\ref{eqcnapart}), we find
that condition (ii) is satisfied in the relevant range of $w$ for
\beq
\frac{1}{s-1} < w < 1 \ . 
\label{wcond}
\eeq
This interval is nonvanishing if $s > 2$.  Thus, for $s > 2$, provided that
conditions (i) and (ii) are satisfied, ${\cal B}_q$ has the form of an open 
self-conjugate arc that crosses the real axis at the point given by 
Eq. (\ref{eqcnapart}), so that in this case,
\beq
q_c = s+1-w(s-1) \quad {\rm for} \ \{G\}=\{C\} \ {\rm and} \ 
 s > 2 \ \ {\rm and} \ \frac{1}{s-1} < w < 1  \ . 
\label{qcdifferent}
\eeq
This self-conjugate arc is concave to the left and ends at the arc endpoints
given by where the ($v=-1$ evaluation of the) expression in the
square root of Eq.  (\ref{lamcnplusminus}) vanishes, namely
\beq
q_e, \ q_e^* = (s+1)(1-w) \pm 2i\sqrt{sw(1-w)} \ . 
\label{qend}
\eeq
This locus does not separate the $q$ plane into different regions.  It is
straightforward to carry out a similar analysis of ${\cal B}_q$ for the FSCP
regime $w > 1$.

\section{Related Topics}

As our results show, one finds a number of intriguing features in the study of
weighted-set vertex coloring of graphs.  There are many further directions of
research in this general area.  One could, for example, use the methods
presented here to calculate $Z(G,q,s,v,w)$ and $Ph(G,q,s,w)$ for other
individual graphs and families of graphs. One could also investigate further
the zeros of these functions in various variables and their accumulation sets
${\cal B}$ for recursive graphs in the limit of infinitely many vertices.  It
would, moreover, be worthwhile to study connections with weighted loop models
\cite{jsloop}.  One could also investigate a different but related type of
graph coloring problem in which the set of colors that one chooses from to
assign to each vertex depends on the vertex.  The unweighted case is called the
list coloring problem in graph theory \cite{woodall}, and it would be useful to
study the weighted-set generalization of list coloring.  We are pursuing these
studies.

\section{Conclusions}

In this paper we have studied the weighted-set graph coloring problems, in
which one assigns $q$ colors to the vertices of a graph such that adjacent
vertices have different colors, with a vertex weighting $w$ that either
disfavors or favors a given set of $s$ colors. In particular, we have analyzed
an associated weighted-set chromatic polynomial $Ph(G,q,s,w)$ and have also
related this to a corresponding Potts model partition function with external 
magnetic fields, $Z(G,q,s,v,w)$.  These functions exhibit a wealth of
interesting properties.  We have proved various general results on these and
illustrated them for particular graphs and families of graphs.  

\begin{acknowledgments}

R.S. thanks Prof. S.-C. Chang for valuable discussions and previous 
collaboration on related work.  This research was partly supported by the grant
NSF-PHY-06-53342.

\end{acknowledgments}

\newpage

\appendix

\vfill
\eject
\end{document}